\documentclass[aps,prd,12pt,nofootinbib,notitlepage]{revtex4}
    \usepackage{graphicx}
    \usepackage{amsmath,amssymb,mathrsfs}
\usepackage{slashed}
\usepackage{epsfig}
\usepackage[font=small,skip=0pt,justification=raggedright]{caption}

\newcounter{fig}   \newcommand{\lbfig}[1]{\refstepcounter{fig}
\label{#1} }
\newcommand{\vphi}{\varphi}

\newcommand{\bea}{\begin{eqnarray}}
\newcommand{\eea}{\end{eqnarray}}
\newcommand{\be}{\begin{equation}}
\newcommand{\ee}{\end{equation}}

\newcommand{\re}[1]{(\ref{#1})}

\usepackage{xcolor}

\usepackage[normalem]{ulem}

\begin{document}

\title{
Hairy dyonic Reissner-Nordstr\"om black holes in an Einstein-Maxwell-Friedberg-Lee-Sirlin type model
}

\author{Jutta Kunz}
\affiliation{Institute of Physics, University of Oldenburg,
Oldenburg D-26111, Germany}
\author{Victor Loiko}
\affiliation{Belarusian State University, Minsk 220004, Belarus}
\author{Yakov Shnir}
\affiliation{BLTP, JINR, Dubna 141980, Moscow Region, Russia\\
Institute of Physics, University of Oldenburg,
Oldenburg D-26111, Germany\\
Hanse-Wissenschaftskolleg, Lehmkuhlenbusch 4, 27733 Delmenhorst, Germany}

\begin{abstract}
    We construct spherically symmetric dyonic black holes in a generalized Maxwell-Friedberg-Lee-Sirlin type model with a complex scalar doublet and a symmetry breaking potential {for the real scalar field}, minimally coupled to Einstein gravity in asymptotically flat space.
    We analyze the properties of the hairy black holes and determine their domain of existence. Our discussion focuses mostly on the case of a long-ranged massless real scalar field.
    Our results indicate that in this case, depending on the coupling constants, the resonant hairy dyonic black holes may bifurcate from Reissner-Nordstr\"om black holes at maximal chemical potential, while the limiting solutions at minimal chemical potential may be related to the Penney solution.
\end{abstract}
\maketitle
\newpage

 \section{Introduction }

  \label{Introduction}

 One of the most interesting developments of the past few decades is related to considerable advances in our understanding of the interplay between black holes and matter fields.
 Circumventing various various no-go theorems in General Relativity - see \cite{Chrusciel:2012jk,Volkov:1998cc,Herdeiro:2015waa} for reviews - it has been found that asymptotically flat black holes can support non-trivial 
 matter fields, forming so-called hairy black holes.
 Well known examples are static black holes with Yang-Mills hair  \cite{Volkov:1998cc,Volkov:1989fi,Bizon:1990sr} and Skyrme hair \cite{Luckock:1986tr,Droz:1991cx}, black holes inside magnetic monopoles \cite{Lee:1991vy,Breitenlohner:1991aa,Breitenlohner:1994di}, dyonic black holes in Einstein-Yang-Mills-Higgs theory \cite{Brihaye:1999nn,Brihaye:1998cm} and black holes with axion hair \cite{Campbell:1991rz}.
 Many of these hairy black holes have been generalized to include spin (see e.g.~\cite{Kleihaus:2000kg,Hartmann:2001ic,Kleihaus:2003sh,Kleihaus:2007vf,Kleihaus:2016rgf}).

Another type of hairy black holes corresponds to Kerr black holes with synchronized complex scalar hair without self-interactions \cite{Hod:2012px,Herdeiro:2014goa,Herdeiro:2014ima}.
Such black holes with stationary linear complex scalar hair are continuously connected to the Kerr black holes.
They bifurcate at the threshold of superradiance \cite{Brito:2015rjv}.
Similar black holes with spinning matter fields were constructed in the non-linear O(3) sigma model \cite{Herdeiro:2018djx} and in Einstein-Proca theory \cite{Herdeiro:2016tmi}.
Black holes of a further type are electrically charged Kerr-Newman black holes with synchronized scalar hair \cite{Benone:2014ssa,Kunz:2019bhm,Herdeiro:2018djx}.

Another class of hairy black holes arises in Einstein-Maxwell theory coupled to a self-interacting charged scalar field, when the superradiance resonance condition \cite{Bekenstein:1973mi,Bekenstein:1998nt} is imposed at the event horizon \cite{Herdeiro:2020xmb,Hong:2020miv} (see also \cite{Hong:2019mcj,Hartnoll:2008vx,Dias:2011tj}).
Notably, such Q-hairy black holes do not emerge from linear clouds around the Reissner-Nordstr\"om (RN) black holes \cite{Hod:2012wmy,Hod:2013nn,Hod:2015hza}.
They feature a gap instead.
On the other hand, Q-hairy black holes are linked to $U(1)$ gauged boson stars, regular localized solutions of the Einstein-Maxwell-scalar theory \cite{Jetzer:1989av,Jetzer:1989us,Kleihaus:2009kr,Pugliese:2013gsa}, which may also possess a regular flat space limit, the $U(1)$ gauged Q-balls.
However, Q-hairy RN black holes may exist in the absence of such a limit, for example, for the dyonic RN black holes with self-interacting $U(1)$ gauged scalar doublet \cite{Herdeiro:2024yqa,Brihaye:2024fjv}.

Flat space Q-balls arise in a variety of models.
In particular, the Friedberg-Lee-Sirlin (FLS) model \cite{Friedberg:1976me,Lee:1991bn} provides an interesting example of a
simple renormalizable two-component scalar field theory with symmetry breaking potential and natural interaction terms \footnote{Recently, the UV-completed modification of the FLS model was considered in \cite{Kim:2024vam}.}.
The FLS model may serve as a prototype to  more realistic theories with symmetry breaking potential like the Standard Model.

The FLS Q-ball solutions can be  smoothly linked to the diverse boson stars in the Einstein-Friedberg-Lee-Sirlin (EFLS) model \cite{Kunz:2019sgn,deSa:2024dhj}.
Further, charged boson stars and RN black holes with Q-hair were constructed recently in the  Einstein-Maxwell-Friedberg-Lee-Sirlin (EMFLS) model \cite{Kunz:2023qfg}.

An interesting aspect of the soliton solutions in the FLS model is that they may also exist in the limiting case of vanishing scalar potential \cite{Levin:2010gp,Loiko:2018mhb,Loiko:2019gwk,Kunz:2021mbm}.
In the EFLS model such configurations correspond to a curious type of hairy black holes with long-ranged scalar hair and a short-ranged massive cloud located on and close to the horizon.
We here investigate the properties of spherically-symmetric solutions of the generalized EMFLS model with a complex scalar multiplet describing asymptotically flat hairy dyonic black holes, putting emphasis on the limit of vanishing potential.

This paper is organized as follows.
In Sec.~II we introduce the model, and the field equations with the stress-energy tensor of the system of interacting fields.
Here we describe the spherically symmetric parametrization of the metric and the matter fields, and we also discuss the physical quantities of interest.
In Sec.~III we present the results of our study with particular emphasis on the limit when the mass of the real scalar field vanishes such that it becomes long-ranged.
We conclude with a discussion of the results and final remarks.

\section{Model}
\subsection{Action and field equations}

We consider Einstein's theory of gravity minimally coupled to an Abelian vector potential $A_\mu$ and a multiplet of $n$ complex, gauged scalar fields $\Phi^{(a)}$ interacting with a real scalar field $\psi$ in an asymptotically flat (3+1)-dimensional spacetime.
The action of the model is
\be
S=\int
    d^4 x \sqrt{- g }\biggl(\frac{R}{4 \alpha^2} + L_m\biggr)\, ,
\label{act}
\ee
where $R$ is the Ricci curvature scalar with respect to the Einstein metric $g_{\mu\nu}$, $g$ is the determinant of the metric tensor, $\alpha^2=4\pi G$, $G$ is Newton's constant, and $L_m$ is the matter field Lagrangian,
\be
L_m=-\frac14 F_{\mu\nu}F^{\mu\nu}-
\sum_{a=1}^n D_{\mu}\Phi^{*(a)}  D^{\mu}\Phi^{(a)}
 - \partial_\mu\psi \partial^\mu\psi - m^2 \psi^2 \sum_{a=1}^n \Phi^{*(a)} \Phi^{(a)}
 -\mu^2 (\psi^2 - v^2)^2 \, .
 \label{FLS}
\ee
The matter field Lagrangian can be considered as an extension of the $U(1)$ gauged two-component FLS model \cite{Friedberg:1976me,Lee:1991bn,Loiko:2019gwk}.
It contains the Maxwell field strength tensor $F_{\mu\nu}=\partial_\mu A_\nu-\partial_\nu A_\mu$, the covariant derivative of the complex multiplet $D_{\mu}\Phi^{(a)}=\nabla_\mu \Phi^{(a)} - iq A_\mu \Phi^{(a)}$ with the gauge coupling constant $q$, and the symmetry breaking potential of the coupled scalar field
\be
U(\Phi, \psi) = m^2 \psi^2 \sum_{a=1}^n \Phi^{*(a)} \Phi^{(a)} + \mu^2 (\psi^2 - v^2)^2 \, .
\label{pot}
\ee
The model \re{act} is invariant under local $U(1)$ gauge transformations
\be
\Phi^{(a)} \to e^{-iq
\chi (x)} \Phi^{(a)},~ A_\mu \to A_\mu + q \partial_\mu \chi(x)\, .
\label{gauge}
\ee

Similarly to the case of the gauged FLS model, the vacuum of the model \re{FLS} corresponds to $D_\mu\Phi^{(a)}=0,~~\partial_\mu\psi = 0$ and $F_{\mu\nu}=0$.
The global minimum of the potential \re{pot} resides at $|\Phi|=0$, $\psi=v$, and the parameter $\mu$ defines the mass of the linearized excitations of the self-interacting real scalar field via $m_\psi = \sqrt 8 \mu v$.
The mass of the components of the complex multiplet $\Phi^{(a)}$ is generated due to the coupling with the real field, $m_\Phi= m v$, for all components.

However, when the real component vanishes, the complex multiplet becomes locally massless for any value of the coupling $m$.
Notably, by choosing a vanishing mass parameter $\mu = 0$, but fixing the vacuum expectation value to $v$, the real massless scalar component $\psi$ becomes long-ranged
\cite{Levin:2010gp,Loiko:2018mhb,Loiko:2019gwk,Kunz:2021mbm}.
The gauge field $A_\mu$ can acquire an effective mass inside the configurations due to the coupling with the complex scalar multiplet.
Towards spatial infinity this effective mass turns to zero, since $|\Phi^{(a)}|=0$ in vacuum.

The complete system of the field equations can be obtained via variation of the action \re{act} with respect to the metric, the gauge potential and the scalar fields.
Then the Einstein equations read
\be
R_{\mu\nu} -\frac12 R g_{\mu\nu} = 8\pi G\left( T_{\mu\nu}^{Em} + T_{\mu\nu}^{\Phi,\psi}\right)\, ,
\label{feq1}
\ee
where the components of the stress-energy tensor of the electromagnetic and the scalar fields are given by
\be
\begin{split}
T_{\mu\nu}^{Em} &=F_\mu^\rho F_{\nu\rho} - \frac14 g_{\mu\nu} F_{\rho\sigma}  F^{\rho\sigma}\,,\\
T_{\mu\nu}^{\Phi,\psi} &=\sum_{a=1}^n\left( D_\mu\Phi^{*(a)} D_\nu\Phi^{(a)} + D_\nu\Phi^{*(a)} D_\mu\Phi^{(a)}\right)
+  \partial_\mu\psi \partial_\nu\psi\\
& - g_{\mu\nu}\left[\frac{g^{\rho\sigma}}{2} \sum_{a=1}^n\left( D_\rho\Phi^{*(a)} D_\sigma\Phi^{(a)} +
D_\sigma\Phi^{*(a)} D_\rho\Phi^{(a)}\right) +
\partial_\rho\psi \partial^\rho\psi +  U(\Phi,\psi)\right] \, .
\end{split}
\label{Teng}
\ee
The gauge field equations are
\be
\partial_\mu(\sqrt{-g} F^{\mu\nu}) = q \sqrt{- g} j^\nu\, ,
\label{feq2}
\ee
where
\be
j_\nu = i\sum_{a=1}^n( D_{\nu}\Phi^{*(a)} \, \Phi^{(a)} - \Phi^{*(a)} D_\nu \Phi^{(a)} )
\label{current}
\ee
is the scalar current associated with the $U(1)$ symmetry.
Finally, the system of the scalar field equations is given by
\bea
\nabla^\mu \nabla_\mu \psi & = & 2\psi \Bigl[ \Bigr. m^2 \sum_{a=1}^n \Phi^{*(a)} \Phi^{(a)}  +2 \mu^2 (v^2-\psi^2))  \Bigl.  \Bigr] \, ,
\label{feqs1}\\
D^\mu D_\mu \Phi^{(a)} & = & m^2 \psi^2 \Phi^{(a)} \, .
\label{feqs2}
\eea

\subsection{Ansatz}

We seek stationary, spherically-symmetric solutions of the system \re{feq1} -  \re{feqs2} describing asymptotically flat hairy dyonic black holes.
For the metric we choose the line element \cite{Kunz:2023qfg,Herdeiro:2024yqa}
\be
ds^2=g_{\mu\nu}dx^\mu dx^\nu= -F_0(r)dt^2 +F_1(r)(dr^2 + r^2 d\Omega^2 ) \, ,
\label{metric}
\ee
where $d\Omega^2= d\theta^2 + \sin^2\theta d\varphi^2$, and the metric functions $F_0$ and $F_1$ depend on the isotropic radial coordinate $r$ only. In such isotropic coordinates the event horizon is characterized
by $F_0=0$ while $F_1$ remains finite there.
It is convenient to implement the following exponential re-parametrization for the metric functions
\be
F_0(r)=\frac{\left(1-\frac{r_H}{r} \right)^2}{\left(1+\frac{r_H}{r} \right)^2} e^{2f_0(r)} \, , \quad
F_1(r)=\left(1+\frac{r_H}{r} \right)^4e^{2f_1(r)}\, ,
\label{metric-hor}
\ee
where $r_H$ is the horizon radius.

For the gauge potential we consider a dyonic ansatz \cite{Herdeiro:2024yqa}
\be
\label{Aans}
A_{\mu} dx^{\mu} =Q_M \cos \theta d\varphi + A_0(r)dt \, ,
\ee
where $Q_M$ is the magnetic charge of the configuration, and $A_0$ is the electric potential.
For an Abelian monopole hidden beyond an event horizon, the Dirac quantization condition holds \cite{Dirac:1931kp}, $qQ_m =N/2$, where $N$ is an integer.
We here consider the complex doublet \re{ansatz-psi} in the presence of a single monopole, $N=1$ \cite{Herdeiro:2024yqa}.

The real scalar field is parameterized as $\psi=X(r)$.
For $n=1$ the model \re{act} is reduced to the usual EMFLS model with a single complex field $\Phi$ \cite{Kunz:2023qfg}.
We here consider the case of a scalar doublet, $n=2$,
since the angular momentum of a single charged scalar field in the Coulomb magnetic field is not zero and such a system would not possess spherical symmetry.
The corresponding ansatz has been proposed in \cite{Gervalle:2022npx,Gervalle:2022vxs,Brihaye:2023vox} and was recently employed in \cite{Herdeiro:2024yqa}.
It reads
\be
\label{ansatz-psi}
\Phi^{(1)} = Y(r) \sin \frac{\theta}{2}~
 e^{i (\frac{\vphi}{2}-\omega t)},~~
 \Phi^{(2)} = Y(r) \cos \frac{\theta}{2}~
 e^{-i( \frac{\vphi}{2}+\omega t)}~,
\ee
thus $\sum_{a} \Phi^{*(a)} \Phi^{(a)}=Y^2$.
We note, that each component of the scalar doublet \re{ansatz-psi} carries a non-zero angular momentum density, however, the total angular momentum of the configuration is zero.

Following our previous study of spherically symmetric boson stars and electrostatic hairy black holes in the EMFLS model \cite{Kunz:2021mbm,Kunz:2023qfg}, we employ for the dyonic hairy black holes the static gauge fixing $A_0(\infty)=0$ and retain the dependence of the solutions on the angular frequency $\omega$.
Then the condition of finiteness of the energy-momentum tensor \re{Teng} and the charge density \re{current} on the event horizon implies the resonance condition between the scalar angular frequency $\omega$ and the value of the electric potential $A_0$ \cite{Herdeiro:2020xmb,Hong:2020miv} (see also \cite{Hong:2019mcj} for the case of clouds)
\be
\omega = q A_0(r_H) \, ,
\label{synchro}
\ee
which needs to be imposed as a boundary condition on the function $A_0$ on the horizon.
We note, that the harmonic time dependence of the scalar doublet can be gauged away.
This changes condition (\ref{synchro}) to $A_0(r_H)=0$, as applied early on for hairy black holes with AdS asymptotics \cite{Hartnoll:2008vx,Dias:2011tj}.
In fact, the resonance condition remains unaffected by the presence of magnetic charge \cite{Herdeiro:2024yqa}.

The full system of equations \re{feq1} - \re{feqs2} is then
solved numerically subject to the boundary conditions
\begin{itemize}
    \item at $r=r_H:\qquad
    \partial_r X(r) = \partial_r Y(r) = \partial_r f_0(r)=\partial_r f_1(r) = 0$ \ ,
    \item at $r=\infty:\qquad
    X(r)=1,\quad Y(r)=0,\qquad f_0(r)=f_1(r)=0$ \, ,
\end{itemize}
which follow from requiring regularity of the fields on the horizon and asymptotic flatness of the metric together with the appropriate vacuum state at spatial infinity.
In addition, the synchronization condition \re{synchro} is imposed.

The usual rescaling of the radial coordinate and the fields allows us to
set $v=1$ and $m=1$, so only four dimensionless model parameters remain,
$q, \mu$, $r_H$ and $\alpha$ \cite{Kunz:2019sgn,Kunz:2021mbm,Kunz:2023qfg,Herdeiro:2024yqa}.

Indeed, two of the parameters of the model \re{act} can be scaled away via transformations of the coordinates and the fields
\be
\psi \to \frac{\psi}{v},\quad \Phi^{(a)}\to
\frac{\Phi^{(a)}}{v}, \quad A_\mu \to \frac{A_\mu}{v}, \quad x_\mu \to mv x_\mu \, ,
\ee
with the remaining rescaled parameters $\alpha \to  v \alpha$, $\mu \to \mu/m$ and $q\to q/m$.

\subsection{Physical quantities}

Expanding the matter fields and the metric functions at the horizon and at spatial infinity allows to evaluate a number of physical observables.
The expansion in the near horizon area reads
\be
\begin{split}
f_0(r)=&f_{00}+f_{02}(r-r_H)^2+{\cal O}(r-r_H)^3 \, ,~~~
f_1(r)=f_{10}+f_{12}(r-r_H)^2+{\cal O}(r-r_H)^3 \, , \\
A_0(r)=&\frac{\omega}{q}+ c_2 (r-r_H)^2+{\cal O}(r-r_H)^3 \, , \\
X(r)=&x_{0}+x_{2}(r-r_H)^2+{\cal O}(r-r_H)^3 \, , \quad
Y(r)=y_{0}+y_{2}(r-r_H)^2+{\cal O}(r-r_H)^3 \, ,
\end{split}
\ee
where $x_0,~y_0,~f_{00},~f_{10}$ and $c_2$ are free parameters.

The relevant horizon quantities include the horizon area $A_H$, the Hawking temperature $T_H$, the horizon mass $M_H$ and the horizon electric charge $Q_H$, where $A_H$ and $T_H$ are given by
\begin{eqnarray}
A_H=64 \pi r_H^2 ~e^{2f_1(r_H)} ,~~
T_H=\frac{1}{64\pi r_H}e^{f_0(r_H)-f_1(r_H)}\, .
\end{eqnarray}
The value of the electromagnetic potential on the event horizon $\mu_{ch}=A_0(r_H)=\omega/q$ defines the chemical potential
of the configuration.

The total ADM mass $M$ and the total electric charge $Q$ of the black holes may be read off from the asymptotic behaviour of metric and gauge field functions for $r\to \infty$
\be
F_0(r) \longrightarrow 1 - \frac{2 M}{r} + {\cal O}\left(\frac{1}{r^2} \right), \quad
A_0(r) \longrightarrow \frac{Q}{r} + {\cal O}\left(\frac{1}{r^2}\right) \, .
\ee
The ADM mass can be represented as the sum of the contributions from the event horizon and the electromagnetic and scalar fields outside the horizon,
\be
M = M_H +  M_V = -\frac{1}{2\alpha^2}\oint\limits_{\Sigma} d\Sigma_{\mu\nu}\nabla^\mu \xi^\nu - \frac{1}{\alpha^2}\int\limits_V d\Sigma_\mu (2T_\nu^\mu \xi ^\nu - T\xi^\mu) \, ,
\ee
with Killing vector field $\xi = \partial_t$.

The electric charge $Q$ also consists of two parts
\be
Q= Q_H+q Q_N \, ,
\ee
where the horizon charge $Q_H$ is given by
\be
Q_H = \frac{1}{4\pi}\oint\limits_{\Sigma} d \Sigma_r F^{0r} =
c_2 r_H^3 e^{f_1(r_H)-f_0(r_H)} \, ,
\ee
and $Q_N$ corresponds to the conserved charge of the Noether current Eq.~(\ref{current})
\be
Q_N =q \int d^3 x \, \sqrt{-g} \, j^0 =
=8\pi \int_{r_H}^\infty dr\, r^2 \frac{F_1^{3/2}}{\sqrt{F_0 }}
 (\omega - q A_0) X^2
\, .
\label{chargeN}
\ee

We note that the presence of a magnetic charge excludes a regular boson star-like limit for $r_H \to 0$ \cite{Herdeiro:2024yqa}.
Nevertheless, it is convenient to introduce the ``hairiness parameter" $h \le 1$ \cite{Herdeiro:2020xmb}
\be
h= 1- \frac{Q_H}{Q}= \frac{q\, Q_N}{Q} \, .
\ee
For dyonic RN black holes $h=0$.
We note, however, that the hairiness parameter does not fully capture the hair of the configurations here, since the real scalar field $\psi$ is not included in the definition.
Clearly, the configurations remain hairy when featuring hair of the real scalar field, while (the hair of) the complex doublet $\Phi$ becomes negligible.\footnote{Alternatively, one might introduce a hairiness parameter $\tilde h=1-\frac{M_{RN}}{M}$, where $M_{RN}$ is the ADM mass of the dyonic RN black hole with the same input parameters $\alpha$, $g$, $r_H$, and $\mu_{ch}$.}

Finally we note, that the solutions satisfy the first law of thermodynamics \cite{Bardeen:1973gs,Gao:2001ut,Herdeiro:2020xmb}. In the grand canonical ensemble the chemical potential $\mu_{ch}$ is fixed and the charge $Q$ is allowed to vary, then, in the units employed,
\be
dM =\frac14 T_H dA_H +\alpha^2 \mu_{ch} dQ \, .
\ee
The corresponding grand potential is
\be
\Omega = M -T_H S -\alpha^2 \mu_{ch} Q \, .
\label{grand}
\ee

\section{Numerical results}

\subsection{Numerical method}

The set of five second order coupled ordinary differential equations for the functions $f_0(r)$, $f_1(r)$, $A_0(r)$, $X(r)$ and $Y(r)$ has been solved numerically subject to the boundary conditions above.
In our numerical calculations we have made use of a sixth-order finite difference scheme, where the system of equations is discretized on a grid with a typical size of 930 points in radial direction.
The emerging system of nonlinear algebraic equations has been solved using the Newton-Raphson scheme.
Calculations have been performed by employing a professional solver \cite{schoen}, with typical errors of order of $10^{-5}$.
To be able to cover the near horizon area with high accuracy also in certain limits we introduced the compactified radial coordinate $x=\frac{r-r_H}{r+C} \in [0,1]$, where $C$ is an arbitrary constant used to adjust the grid according to the contraction of the solutions.

Since the model still possesses a large space of parameters, $q$, $\mu$, $r_H$ and $\alpha$, we here fix the value of the gauge coupling to $q=0.1$.
Thus the variation of the angular frequency $\omega$ is directly fixed by the change of the chemical potential $\mu_{ch}=\omega/q$.
Following previous works \cite{Loiko:2018mhb,Kunz:2019sgn,Loiko:2019gwk,Loiko:2020htk,Kunz:2023qfg,Kunz:2021mbm}, we further restrict our analysis to two particular values of the mass parameter $\mu$ of the real scalar field, $\mu^2 = 0.25$ and $\mu = 0$.
The latter value corresponds to the limiting case of vanishing potential.
We note that setting $\mu=0$ changes the asymptotic behavior of the then massless real scalar field.
Thus the function $X$ decays as \cite{Levin:2010gp}
\be
X(r)=1-\frac{D}{r} + {\cal O}\left(\frac{1}{r^2}\right) \, .
\label{massless_tail}
\ee
This is a long-range Coulomb-like asymptotic fall-off with a scalar charge $D$.

\subsection{Dyonic Reissner-Nordstr\"om black holes with resonant scalar hair: Finite mass \boldmath $\mu$ \unboldmath } 

\begin{figure}[t!]
\begin{center}
\includegraphics[height=.33\textheight,  angle =-90]{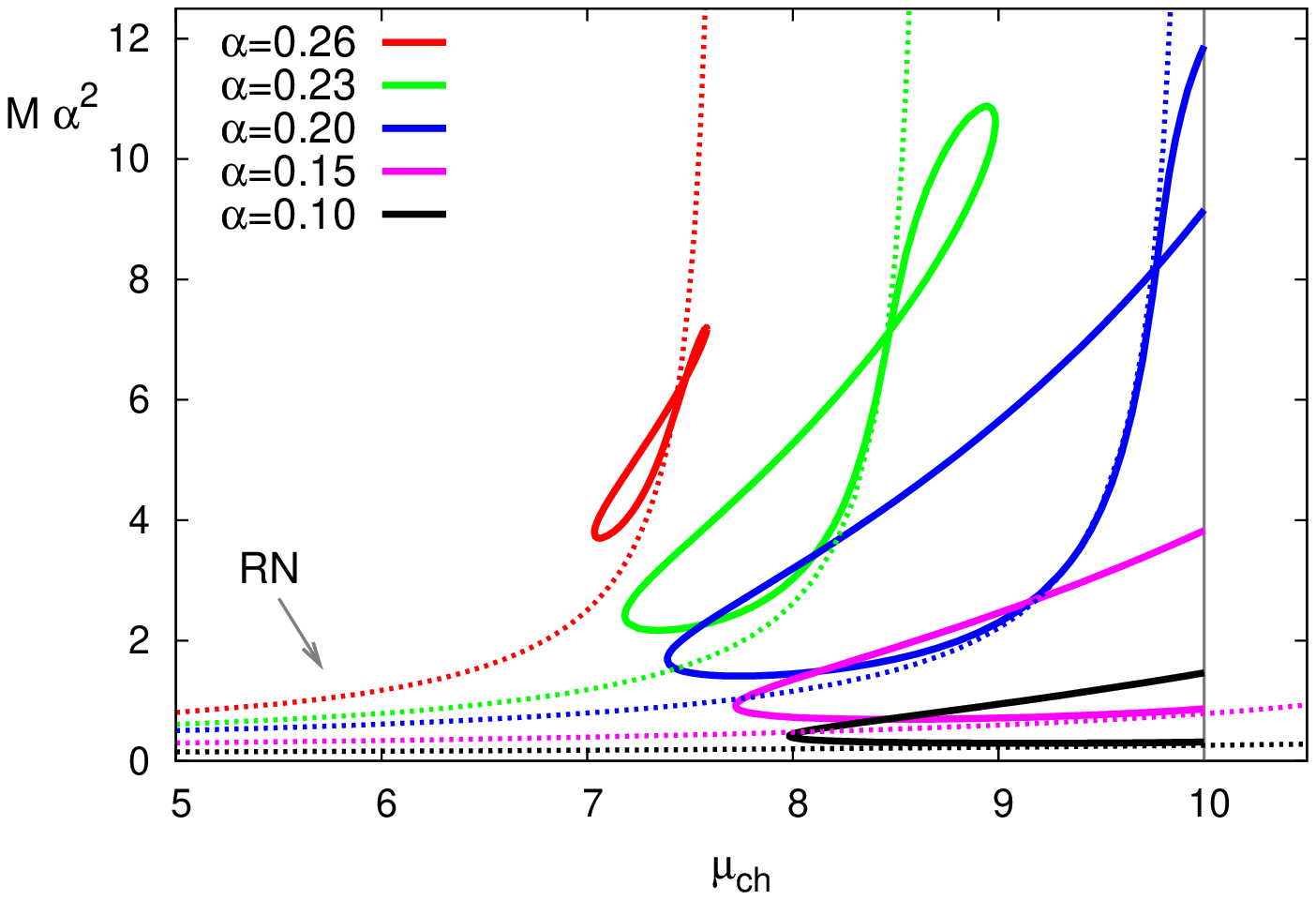}
\includegraphics[height=.33\textheight,  angle =-90]{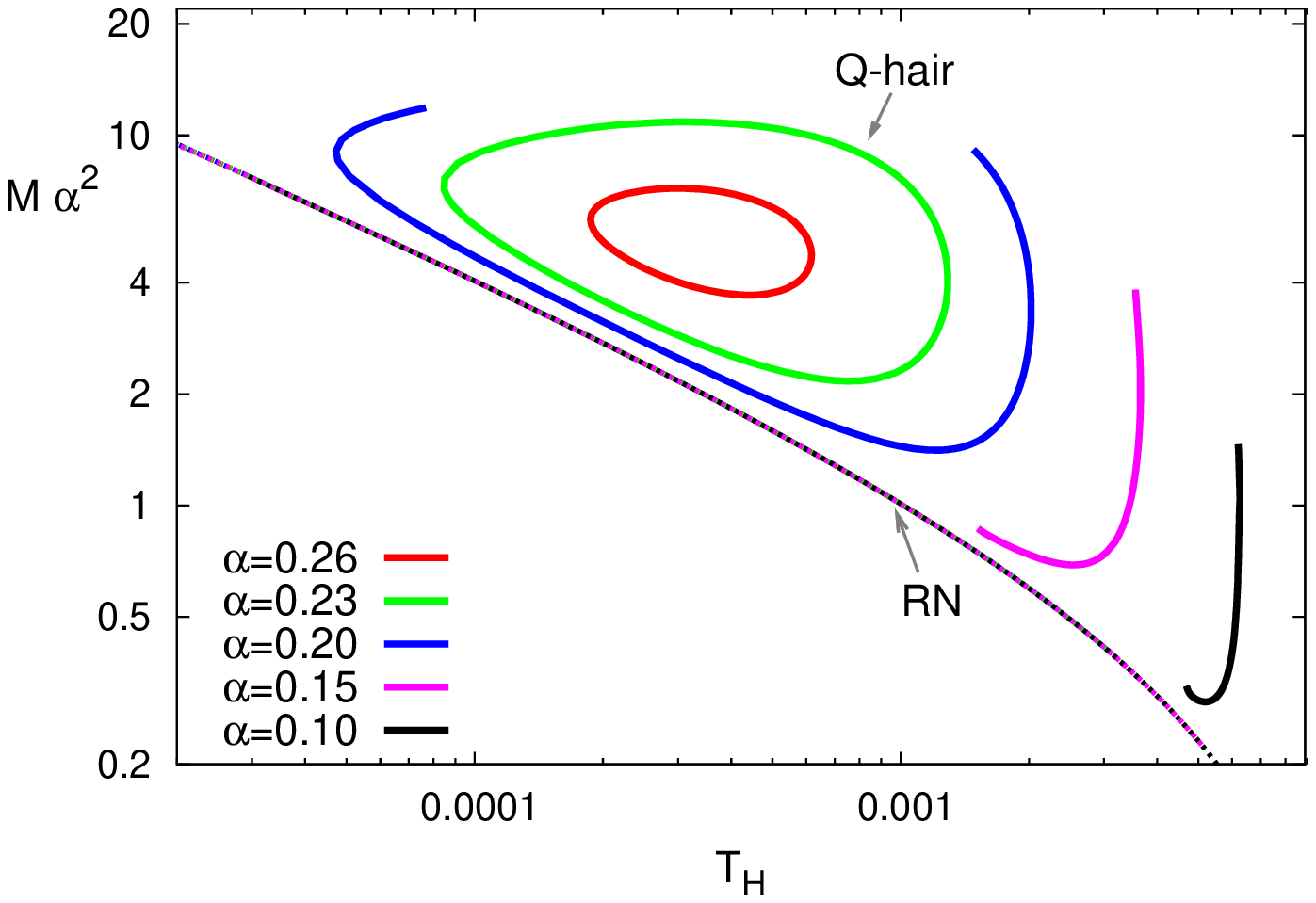}
\includegraphics[height=.33\textheight,  angle =-90]{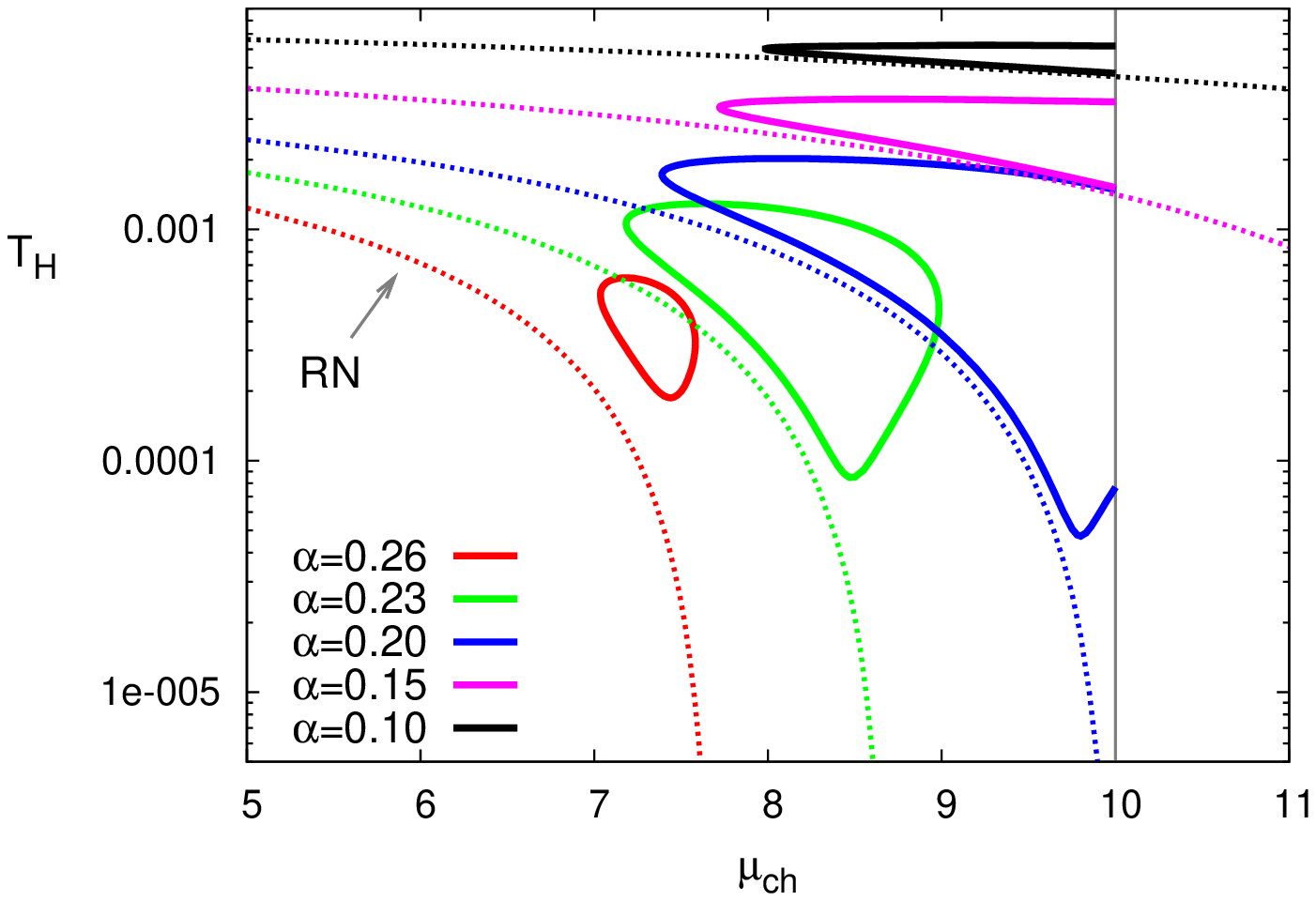}
\includegraphics[height=.33\textheight,  angle =-90]{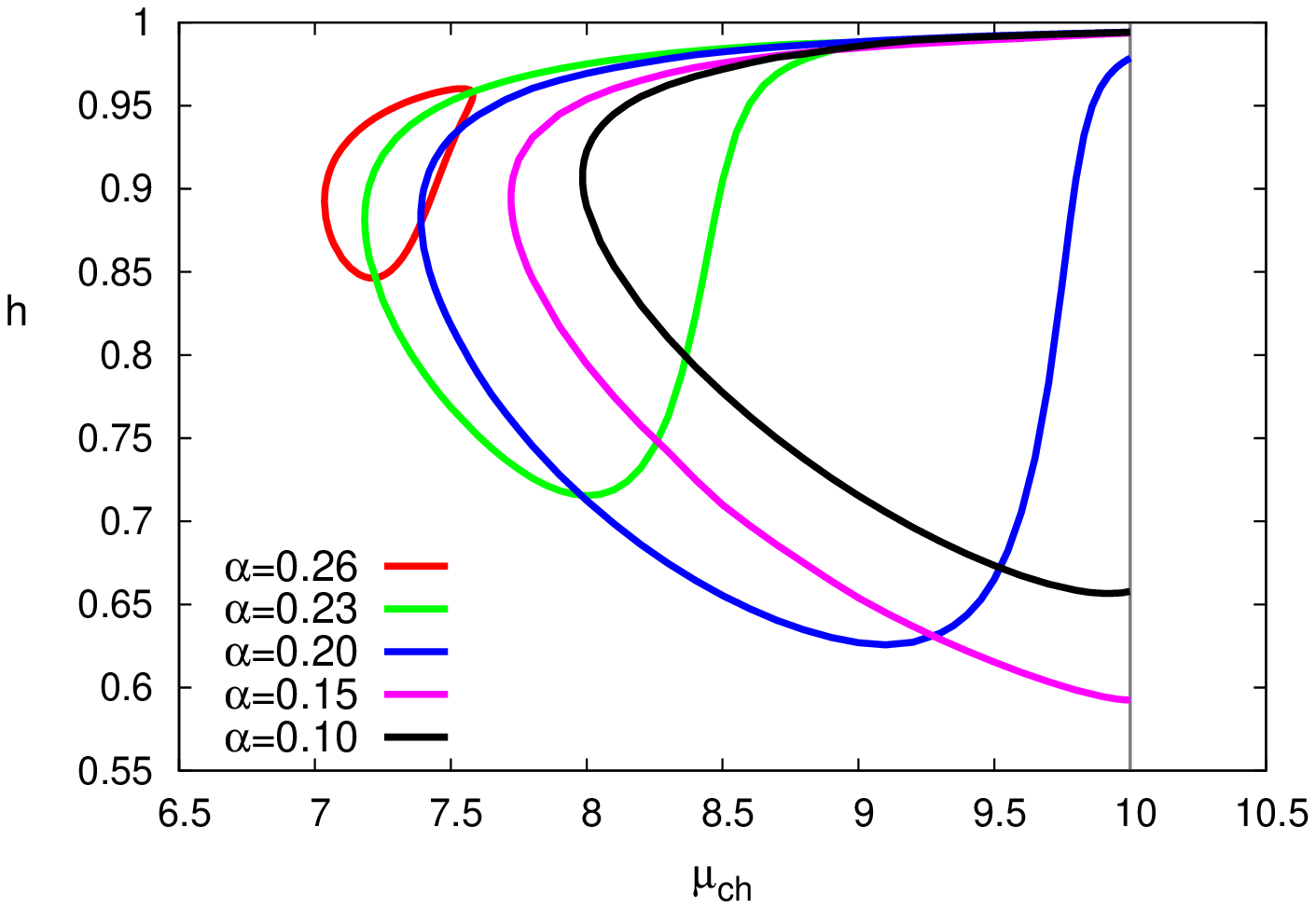}
\end{center}
\caption{\small
Dyonic RN black holes with resonant Q-hair:
Mass $M$ vs chemical potential $\mu_{ch}$ (upper left) and vs Hawking temperature $T_H$ (upper right);
Hawking temperature $T_H$ (lower left)
and hairiness parameter $h$ (lower right) vs chemical potential $\mu_{ch}$ for a set of values of the gravitational coupling $\alpha$, for horizon radius $r_H=0.1$, gauge coupling $g=0.1$ and mass parameter $\mu^2=0.25$.
For comparison the corresponding dyonic RN properties are also shown (dotted curves).
}
    \lbfig{fig1}
\end{figure}

For a finite mass parameter $\mu$ the numerical calculations reveal a pattern that is qualitatively similar to the pattern observed for the spherically symmetric electrically charged black holes with resonant EMFLS Q-hair \cite{Kunz:2023qfg} and also for the RN dyonic black holes in the model with scalar multiplets and polynomial potential \cite{Herdeiro:2024yqa}.

In order to present our results we first fix a particular value of the horizon radius $r_H$ and consider variations of the chemical potential $\mu_{ch}$ for a set of values of the gravitational coupling $\alpha$.
The solutions exist within a restricted interval of values of the chemical potential $\mu_{ch} \in [\mu_{ch}^{(min)},~\mu_{ch}^{(max)}]$.
As the gravitational coupling $\alpha$ remains relatively weak, $\alpha \lesssim 0.21$, the maximal value $\mu_{ch}^{(max)}$ corresponds to the mass of the excitations of the scalar field, where our choice of parameters yields $\mu_{ch}^{(max)}=10$.

In Fig.~\ref{fig1}, upper left plot, we present the total mass $M$ of the charged black holes with resonant Q-hair versus their chemical potential $\mu_{ch}$ for a set of values of the gravitational coupling $\alpha$ and fixed isotropic horizon radius $r_H=0.1$ and gauge coupling $g=0.1$.
Since the event horizon is endowed with the electric charge $Q_H$ an electrostatic repulsive force appears acting on the charged $Q$-cloud surrounding the black hole.
Consequently, black holes with resonant charged hair exist because electromagnetic repulsion and gravitational attraction balance one another.

For relatively weak gravitational coupling $\alpha$, the pattern of dynamical evolution of the Q-hairy black holes in the model \re{act} is not very different from the case of regular charged solitonic solutions, discussed in  \cite{Kleihaus:2009kr,Pugliese:2013gsa,Kunz:2021mbm}.
As the chemical potential is decreased below the upper critical value, a fundamental branch of the charged hairy black holes arises.
This branch extends backward up to a minimal value of the chemical potential $\mu_{ch}^{(min)}$, where a bifurcation occurs and a second, forward branch is  encountered.
Along the fundamental branch the electromagnetic energy remains much smaller than the total energy of the configuration, and the mass of the dyonic black holes with resonant scalar hair is slightly higher than the mass of the corresponding RN black holes, while getting closer to it with increasing $\mu_{ch}$.

On the upper (electrostatic) branch the mass of the Q-hairy black holes is much higher.
Here the corresponding hairiness parameter $h$ is significantly larger than on the lower (scalar) branch, as seen in Fig.~\ref{fig1} (lower right).
The electrostatic repulsion becomes stronger than the gravitational attraction, the charged cloud is expanding rapidly until the chemical potential attains its threshold value.

This pattern changes as the gravitational coupling increases.
The stronger gravitational attraction then prevents an inflationary expansion of the Q-clouds with increasing mass, and the upper critical value of the chemical potential is shifted down from the threshold, as seen in Fig.~\ref{fig1}.
Notably, the dyonic black holes with resonant hairs do not develop a spiralling pattern, which is typical for boson stars, because the electrostatic repulsive force is acting on the charged Q-clouds from the black hole horizon with horizon charge $Q_H$.

Interestingly, there is a particular value of the chemical potential, where the competing forces of gravitational attraction and electrostatic repulsion just balance each other.
At this value the mass of the hairy black holes coincides with the mass of the dyonic RN black holes, as seen in Fig.~\ref{fig1} (upper left).
The electrostatic repulsion affects the lower bound of the range of values of the chemical potential $\mu_{ch}$, as seen in Fig.~\ref{fig1}.
The allowed range of values of the chemical potential decreases as the gravitational coupling increases, and hairy black holes cease to exist at some critical value of $\alpha$,  $\alpha_{\rm max}\sim 0.29$.

\begin{figure}[t!]
\begin{center}
\includegraphics[height=.33\textheight,  angle =-90]{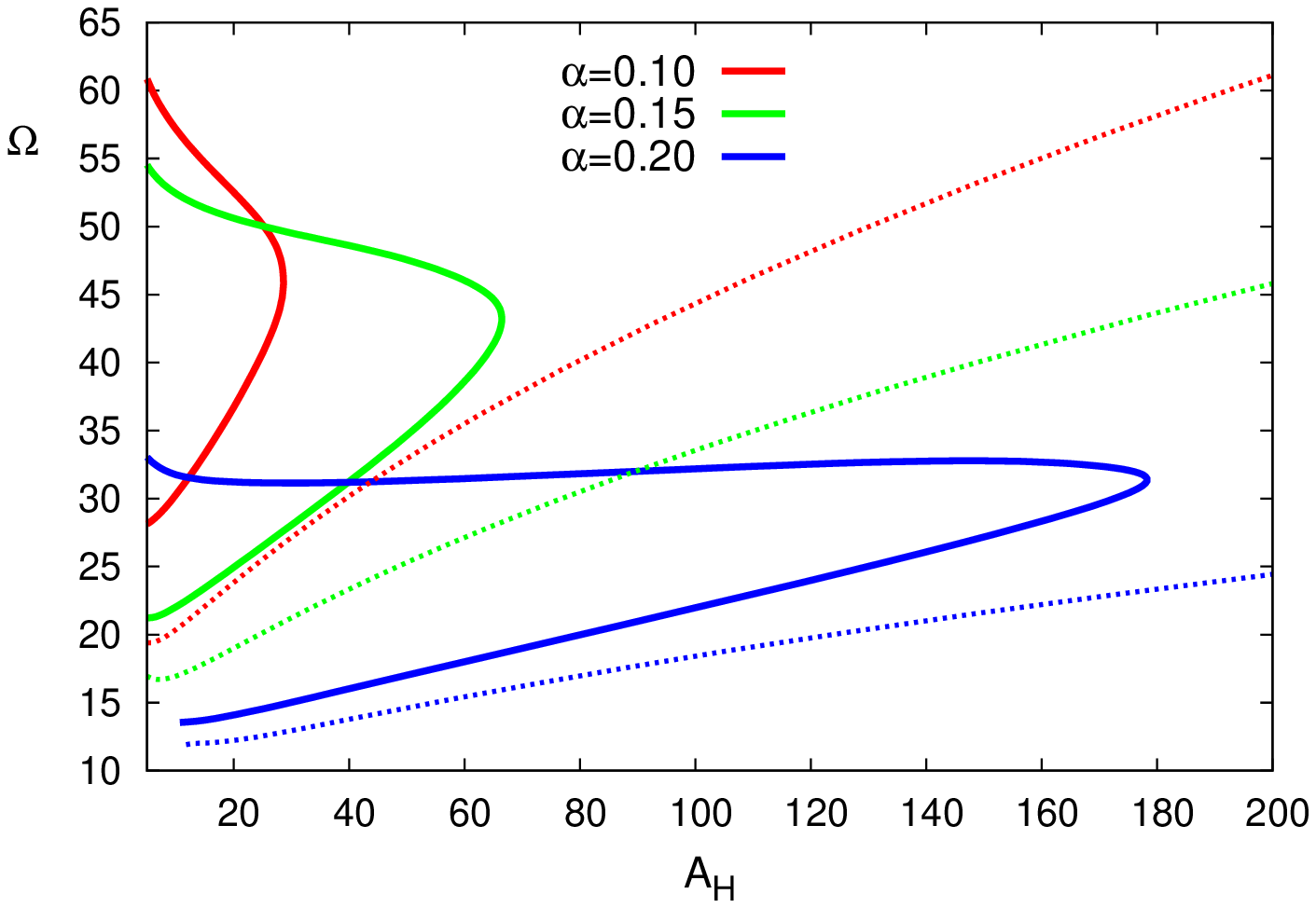}
\includegraphics[height=.33\textheight,  angle =-90]{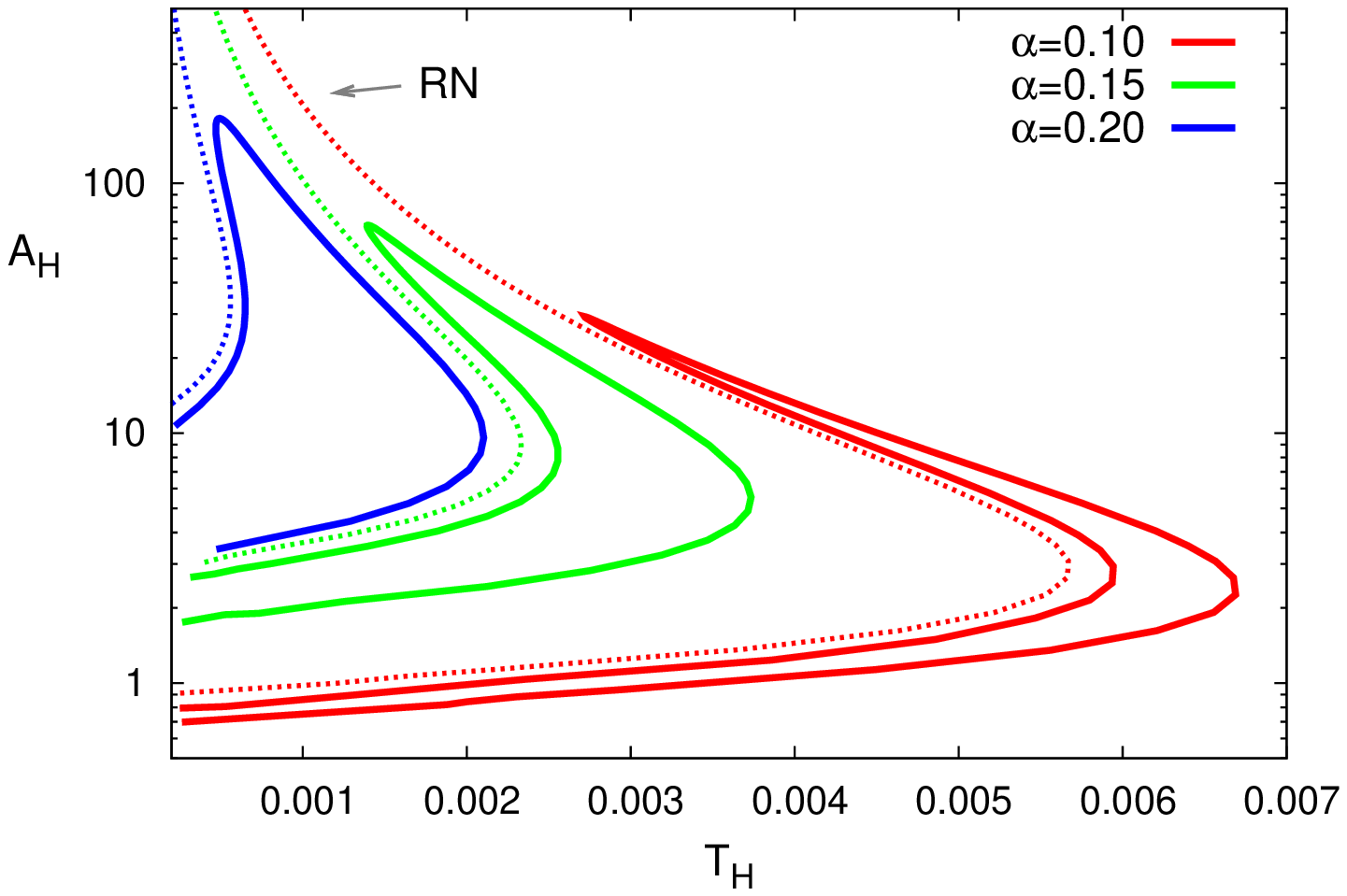}
\end{center}
\caption{\small
Dyonic RN black holes with resonant Q-hair:
Grand potential $\Omega$ (left) and Hawking temperature $T_H$ (right) vs horizon area $A_H$ for
fixed chemical potential $\mu_{ch}=8.5$ and a set of values of the gravitational coupling
$\alpha$, for gauge coupling $g=0.1$ and mass parameter $\mu^2=0.25$.
For comparison the corresponding dyonic RN properties are also shown (dotted curves).
}
   \lbfig{fig2}
\end{figure}

We display in Fig.~\ref{fig2} the grand potential $\Omega=M$, Eq.~(\ref{grand}), and the Hawking temperature $T_H$ of hairy dyonic black holes in a grand canonical ensemble, i.e., for a fixed value of the chemical potential, $\mu_{ch}=8.5$, versus the horizon area $A_H$ for a set of values of the gravitational coupling $\alpha$ and for gauge coupling $q=0.1$ and mass parameter $\mu^2= 0.25$.
Our results indicate that the grand potential of the hairy black holes is always higher than that of RN black holes at fixed chemical potential for a given entropy or temperature, rendering the RN black holes the thermodynamically preferred black holes.
The double branch structure, which was observed in similar cases \cite{Kunz:2023qfg,Herdeiro:2024yqa} still persists.
Both branches bifurcate at a maximal value of the horizon area $A_H$.
The branches originate from two different $T_H\to 0$ critical solutions possessing nonzero values of the horizon area $A_H$, while their mass $M$ and electric charge $Q$ remain finite.

The Hawking temperature $T_H$ of the hairy dyonic black holes is always higher than the one of the electrovacuum dyonic RN black holes, as seen in Figs.~\ref{fig1} and \ref{fig2}.
Note that the dyonic black holes with resonant Q-hair never reach the RN limit, although the lower branch of solutions follows very closely the corresponding dyonic RN black holes with the same values of $Q_m$, $\mu_{ch}$ and $r_H$.
Indeed, the hairiness parameter $h$ never vanishes and it also never approaches unity, as seen in Fig.~\ref{fig1} (lower right).

\subsection{Dyonic Reissner-Nordstr\"om black holes with resonant scalar hair: \\ massless limit \boldmath $\mu=0$ \unboldmath}

\begin{figure}[t!]
\begin{center}
\includegraphics[height=.33\textheight,  angle =-90]{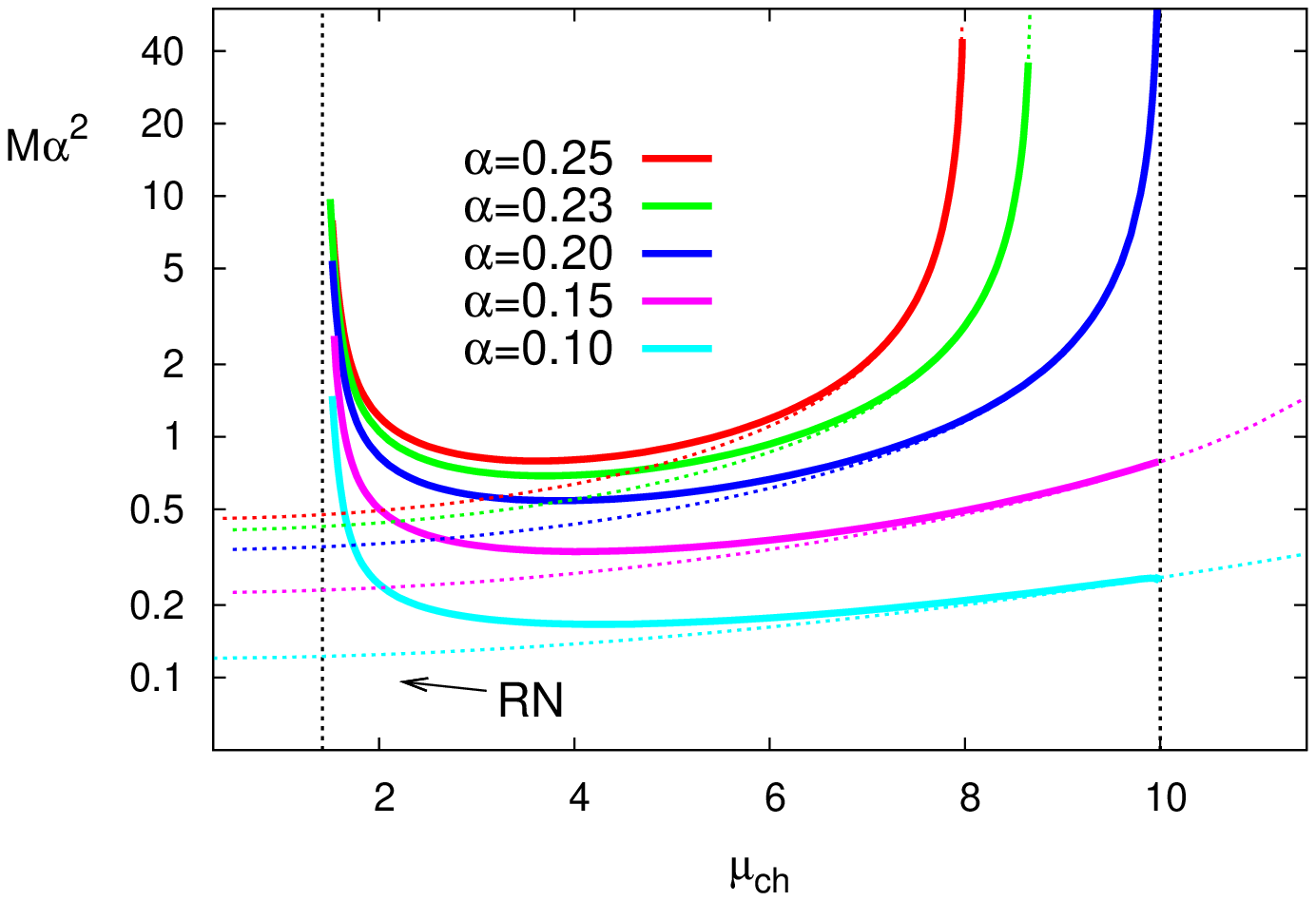}
\includegraphics[height=.33\textheight,  angle =-90]{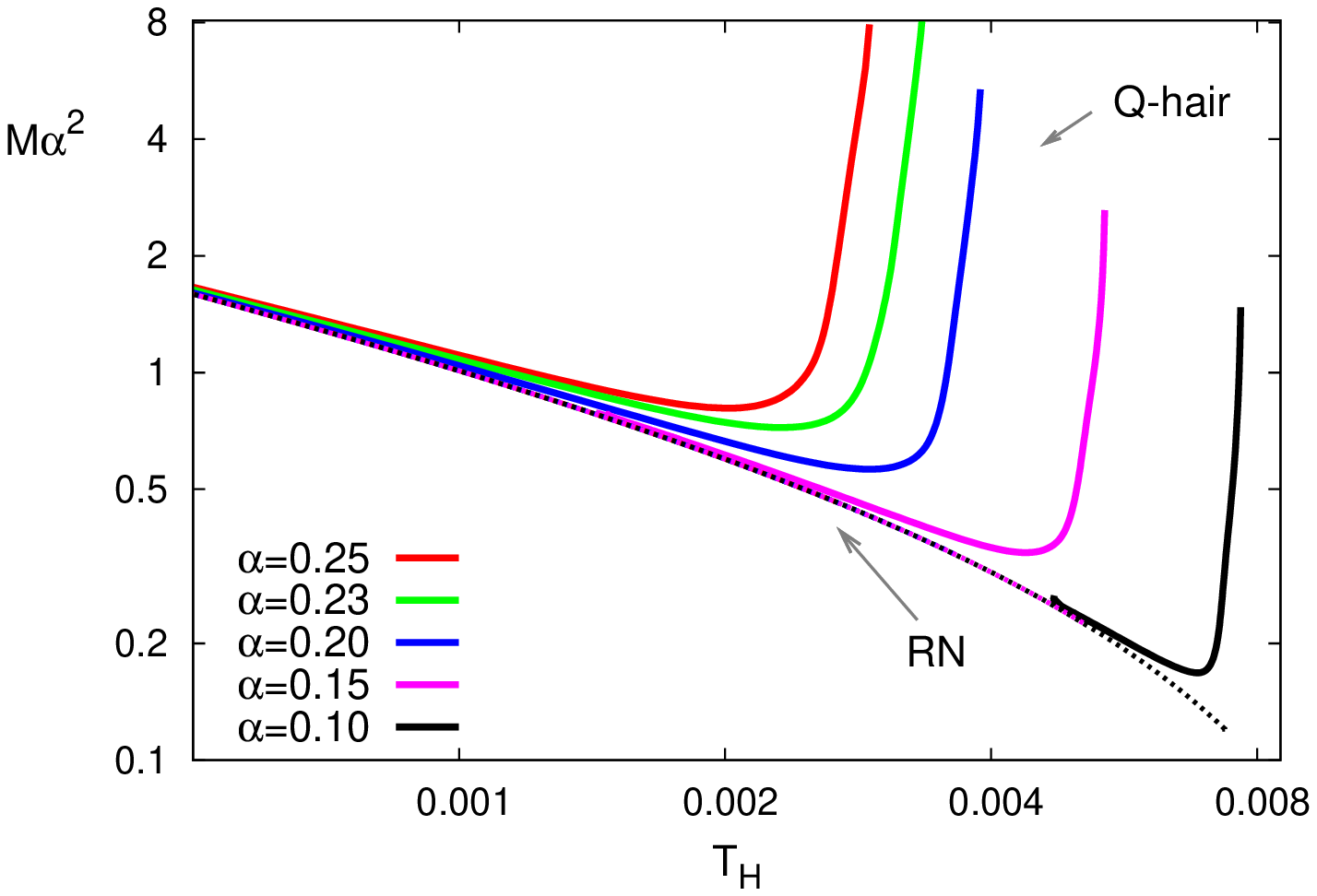}
\includegraphics[height=.33\textheight,  angle =-90]{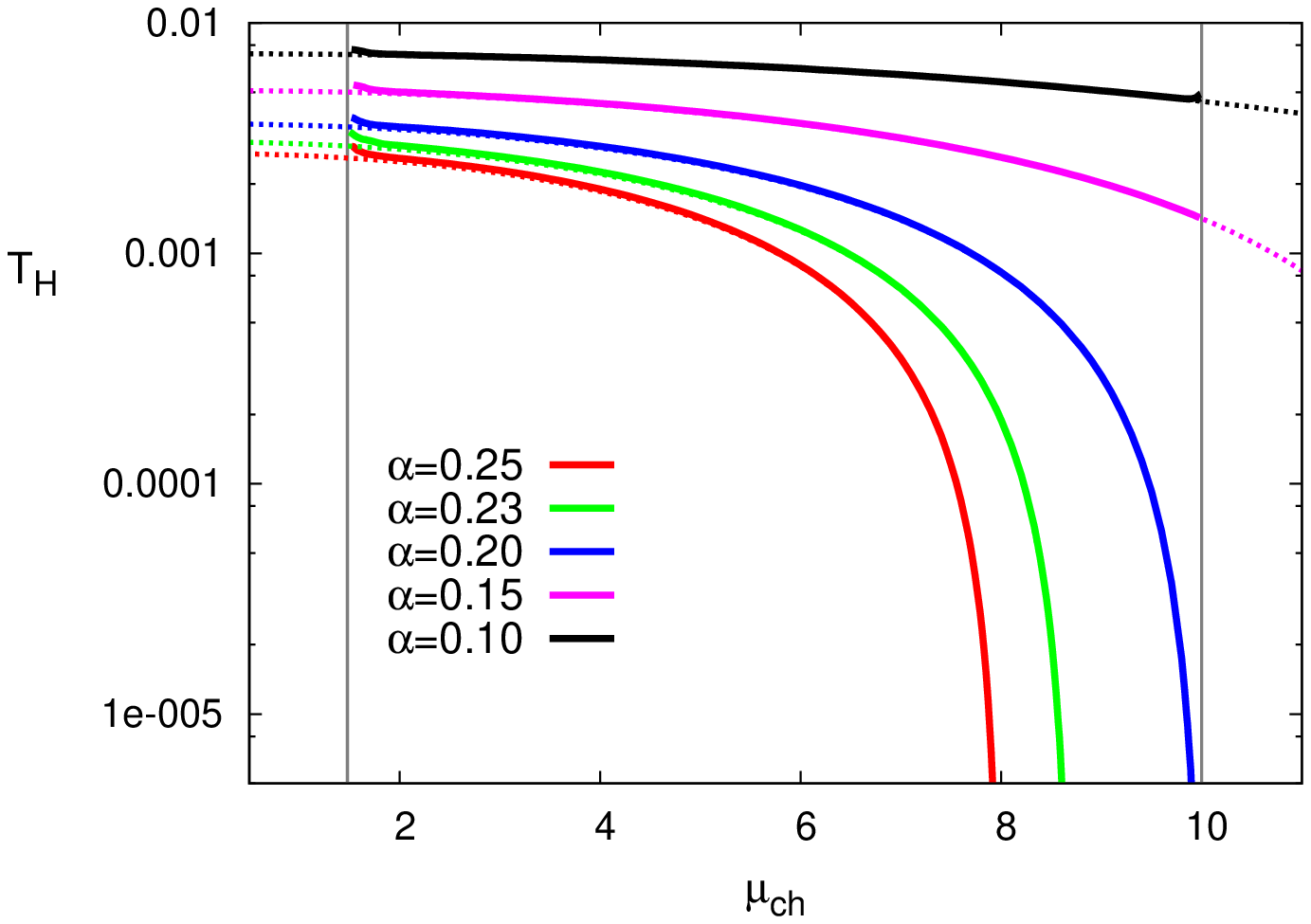}
\includegraphics[height=.33\textheight,  angle =-90]{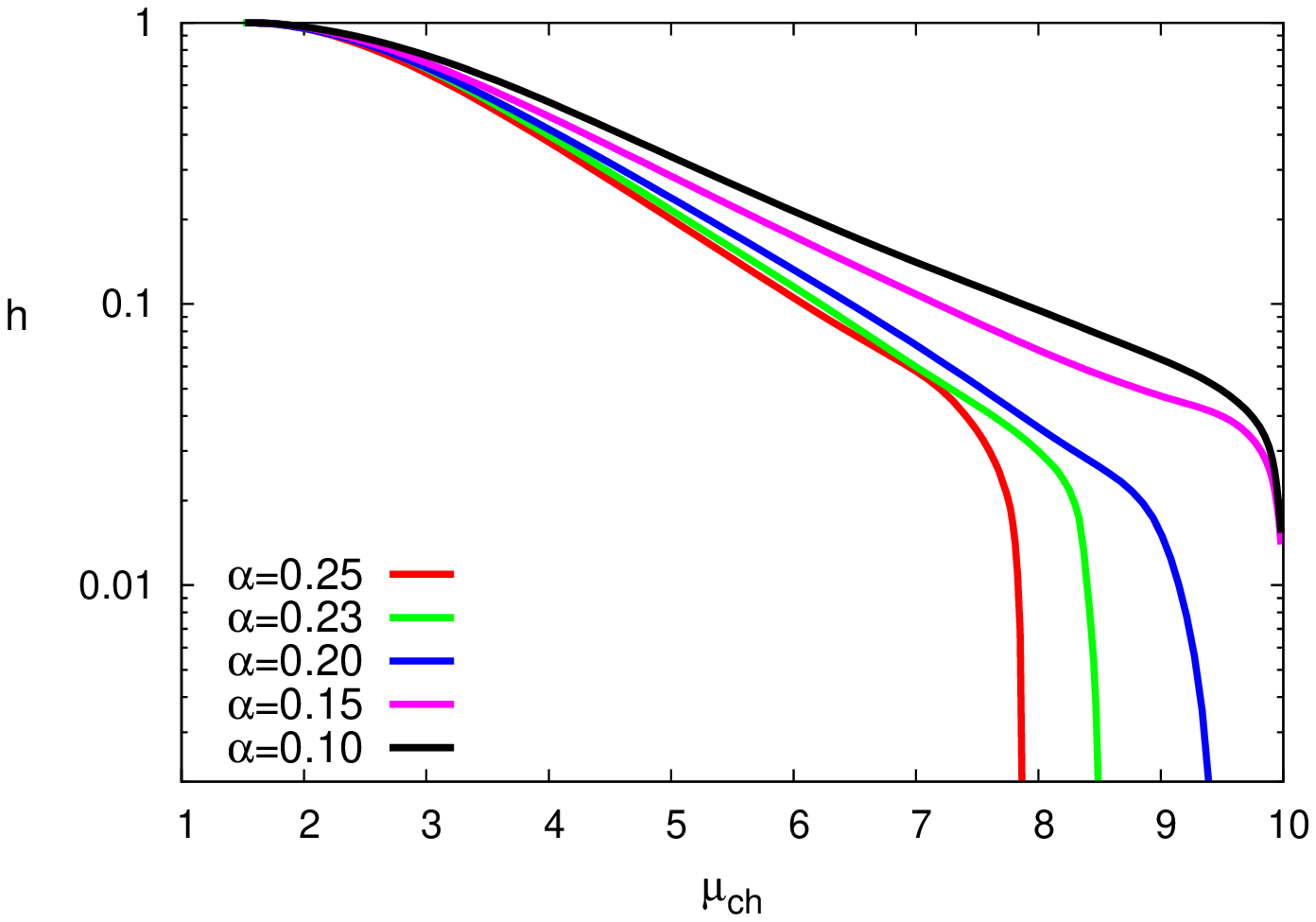}
\includegraphics[height=.33\textheight,  angle =-90]{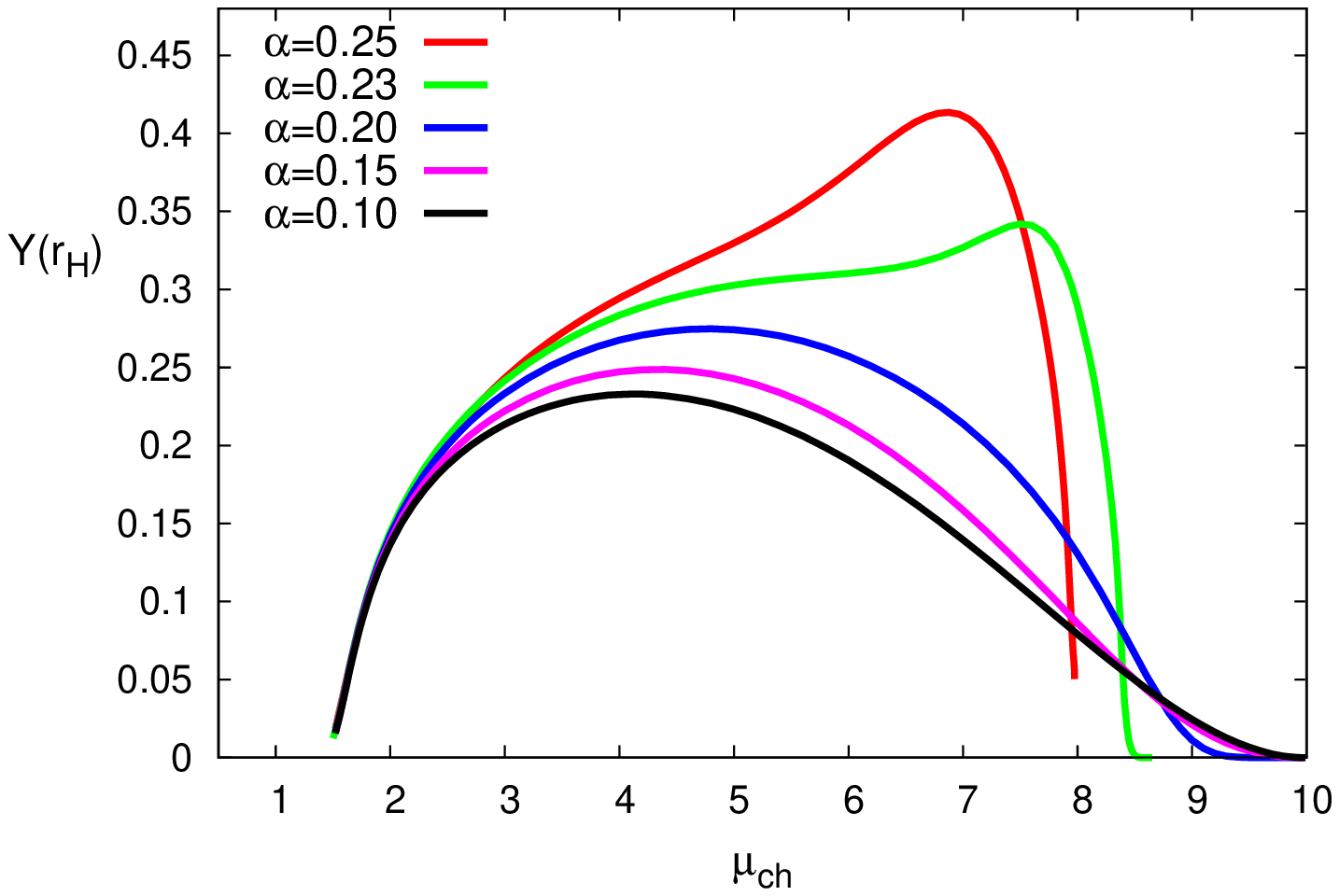}
\includegraphics[height=.33\textheight,  angle =-90]{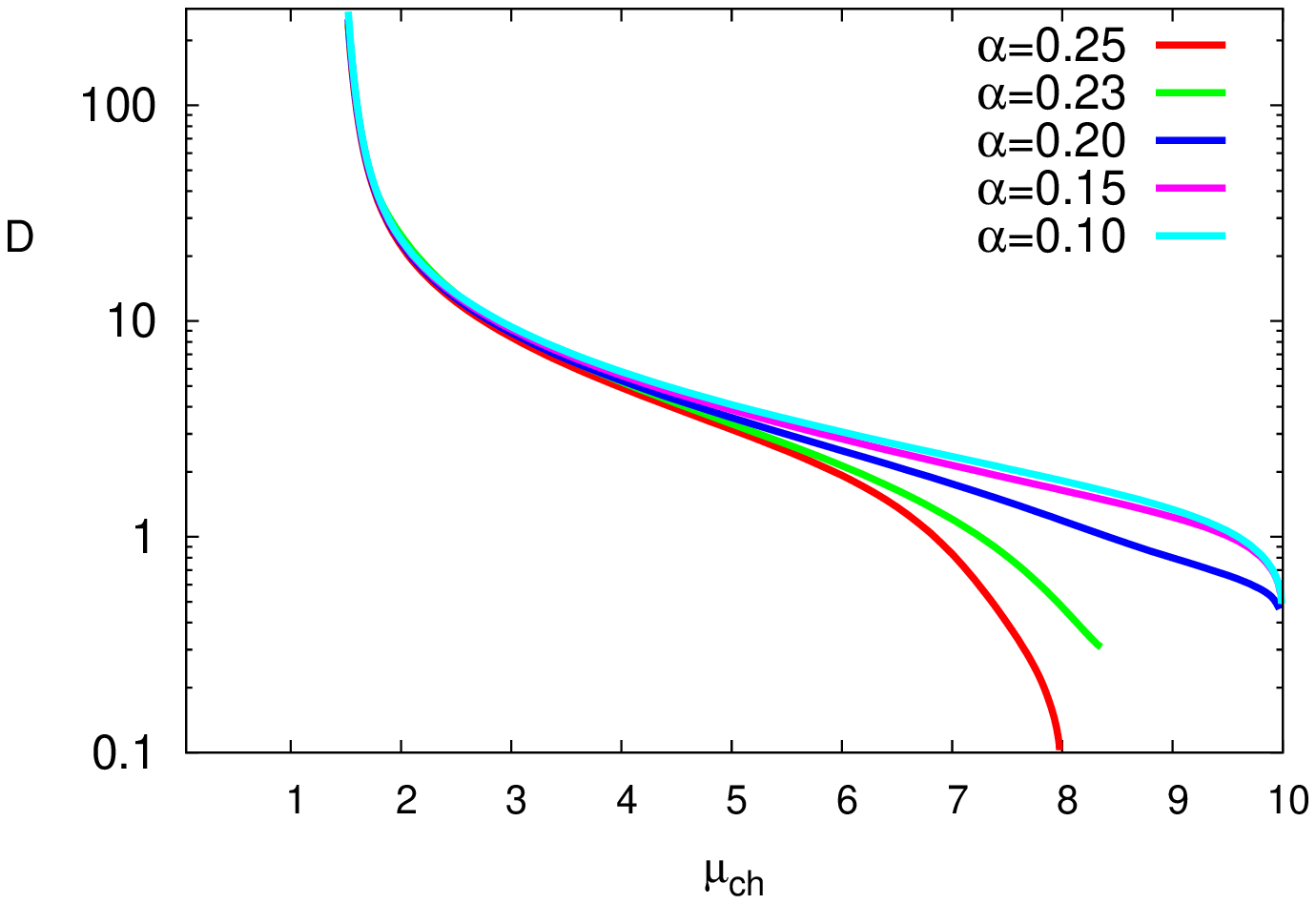}
\end{center}
\caption{\small
Dyonic RN black holes with resonant Q-hair in the massless ($\mu=0$) limit:
Mass $M$ vs chemical potential $\mu_{ch}$ (upper left) and vs Hawking temperature $T_H$ (upper right);
Hawking temperature $T_H$ (middle left), hairiness parameter $h$ (middle right), horizon value of profile function $Y(r_H)$ (lower left) and scalar charge $D$ (lower right) vs chemical potential $\mu_{ch}$ for a set of values of the gravitational coupling $\alpha$, for horizon radius $r_H=0.1$, and gauge coupling $g=0.1$.
For comparison the corresponding dyonic RN properties are also shown (dotted curves).
}
\lbfig{fig3}
\end{figure}
\begin{figure}[t!]
\begin{center}
\includegraphics[height=.33\textheight,  angle =-90]{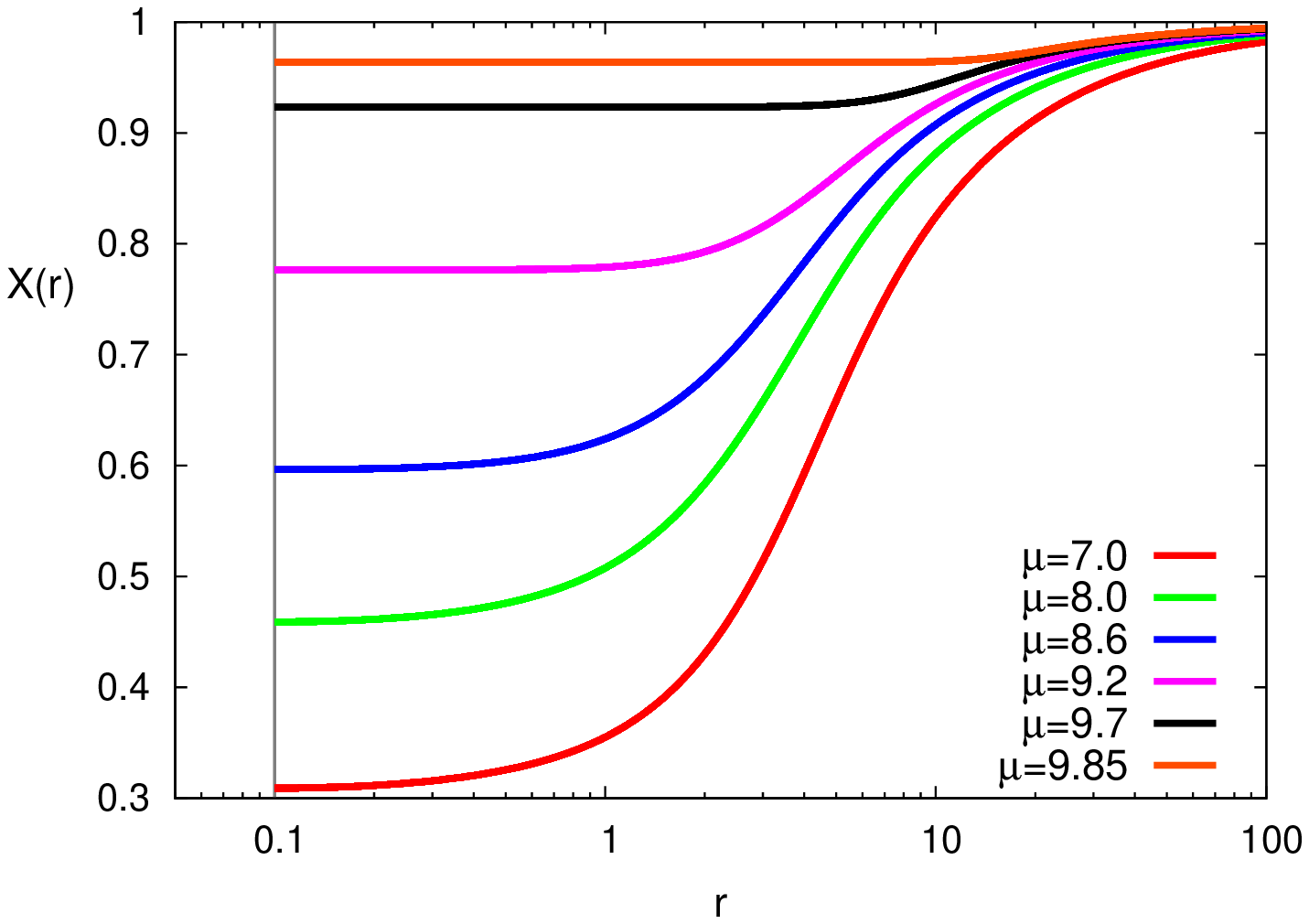}
\includegraphics[height=.33\textheight,  angle =-90]{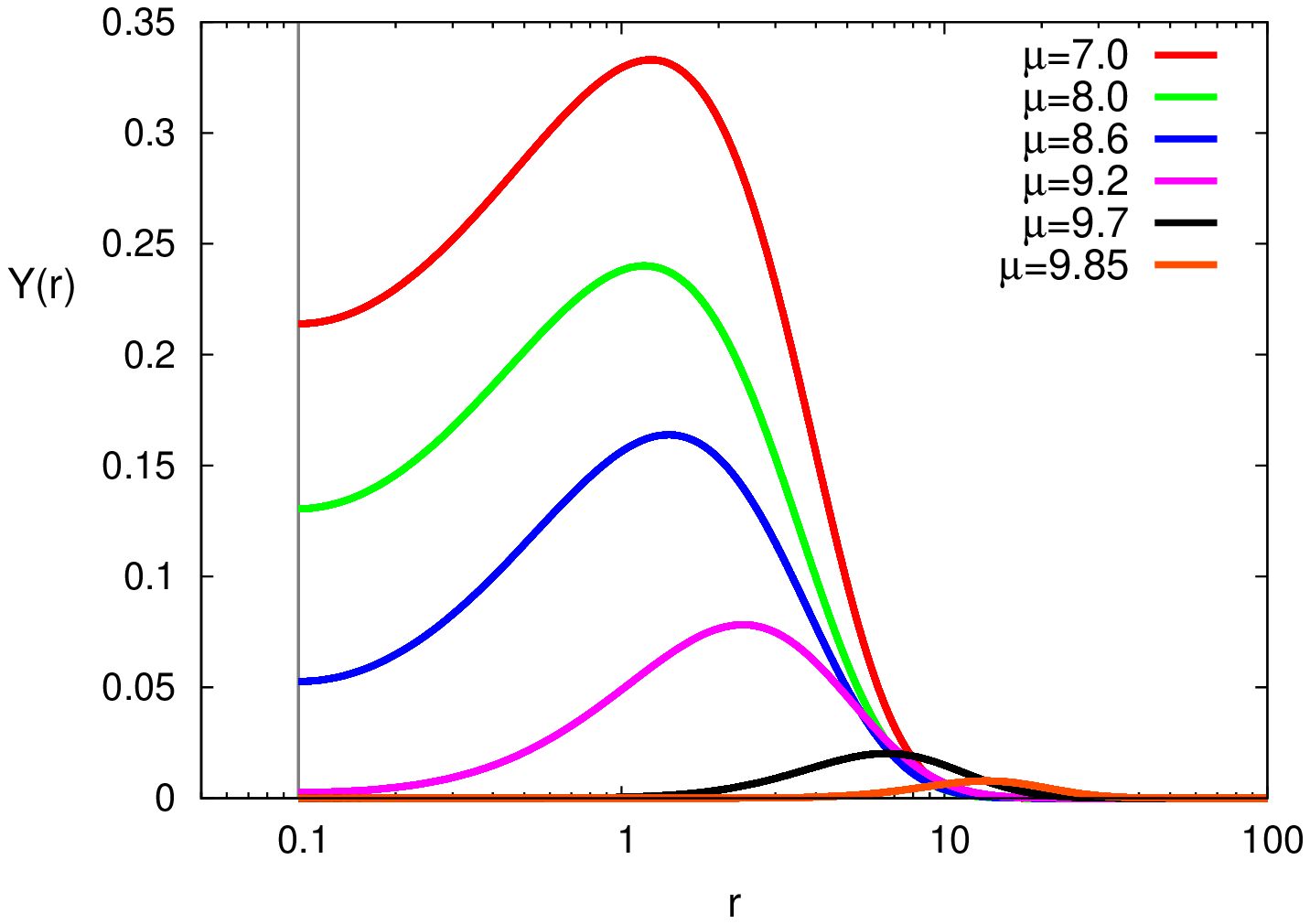}
\end{center}
\caption{\small
Dyonic RN black holes  with resonant Q-hair in the massless ($\mu=0$) limit:
Profile functions of the scalars $X(r)$ (left) and $Y(r)$ (right) for a set of values of the chemical potential $\mu_{ch}$ for gravitational coupling $\alpha=0.20$, horizon radius $r_H=0.1$, and gauge coupling $g=0.1$.
}
    \lbfig{fig4}
\end{figure}

Remarkably, localized solutions of the extended EMFLS-type model \re{act} with potential \re{pot} exist in the limit of vanishing potential $\mu=0$, when the vacuum expectation value of the real component $\psi$ is kept non-zero \cite{Levin:2010gp,Loiko:2018mhb,Loiko:2019gwk,Kunz:2021mbm}, and the system possesses a residual scaling symmetry.
Setting the mass parameter $\mu=0$ changes the asymptotic behavior of this component, leading to its Coulomb-like decay \re{massless_tail}.
The scalar charge $D$, which corresponds to the Coulomb asymptotic tail of the massless scalar field $\psi$ is not constant.
Instead, analogous to the electric charge $Q$ and the ADM mass $M$ of the configurations, also the scalar charge $D$ depends on the parameters of the system.

Importantly, the features of the hairy dyonic RN black holes are related to the delicate force balance between gravitational attraction, electrostatic repulsion and the short and long range scalar interactions.
The appearance of a massless scalar component changes the balance of forces dramatically \cite{Loiko:2018mhb,Kunz:2019sgn,Kunz:2021mbm,Kunz:2023qfg}.
We note that the complex fields $\Phi^{(1)}$ and $\Phi^{(2)}$ remain massive.
But when the long-range scalar field $\psi$ approaches zero within some region around the center of the configuration, the complex fields become locally massless in this region.

To investigate the properties of the hairy dyonic black holes in the limit $\mu=0$, we first keep the horizon radius $r_H$ and the gauge coupling $q$ fixed, and consider the dependence of the solutions on the chemical potential $\mu_{ch}$ for several values of the gravitational coupling.
In Fig.~\ref{fig3} (upper plots) we exhibit the ADM mass $M$ of the hairy black holes versus the chemical potential $\mu_{ch}$ and the Hawking temperature $T_H$.
For comparison we also include the corresponding properties of the dyonic RN black holes, when suitable.

Notably, in the massless limit $\mu=0$ there is only a single branch of solutions starting from the maximal critical value of the chemical potential $\mu_{ch}^{(max)}$ and extending down to some non-zero minimal value $\mu_{ch}^{(min)}$.
The resulting pattern crucially depends on the strength of the gravitational coupling $\alpha$.
Figure \ref{fig3} (upper left) shows that for relatively weak coupling, $\alpha \lesssim 0.20$, the branch of Q-hairy dyonic black holes emerges smoothly in the background of a dyonic RN black hole at the maximal threshold value of the chemical potential $\mu_{ch}^{(max)} = 10$.
Indeed, also the other observables displayed in Fig.~\ref{fig3} versus the chemical potential, the hairiness parameter (middle right), the horizon value of the profile function $Y(r_H)$ (bottom left) and the scalar charge $D$ (bottom right), indicate that a small non-linear spherical charged Q-shell is formed in a distant region far from the event horizon.

The field of the complex doublet is zero between the shell and the horizon.
This is illustrated in Fig.~\ref{fig4}, which demonstrates how the components of the scalar fields emerge from the dyonic RN vacuum, as the chemical potential $\mu_{ch}$ decreases below the maximal threshold value.
The shell is stabilized by the competing forces of long range scalar interaction, electrostatic repulsion and gravitational attraction.
As the chemical potential decreases, the charged shell moves towards the horizon and the horizon value of the profile function $Y(r_H)$ increases from zero, as seen in Fig.~\ref{fig3} (lower left).
Thus, a Q-cloud is formed around the RN black hole, as seen in Fig.~\ref{fig4}.

\begin{figure}[t!]
\begin{center}
\includegraphics[height=.34\textheight,  angle =-90]{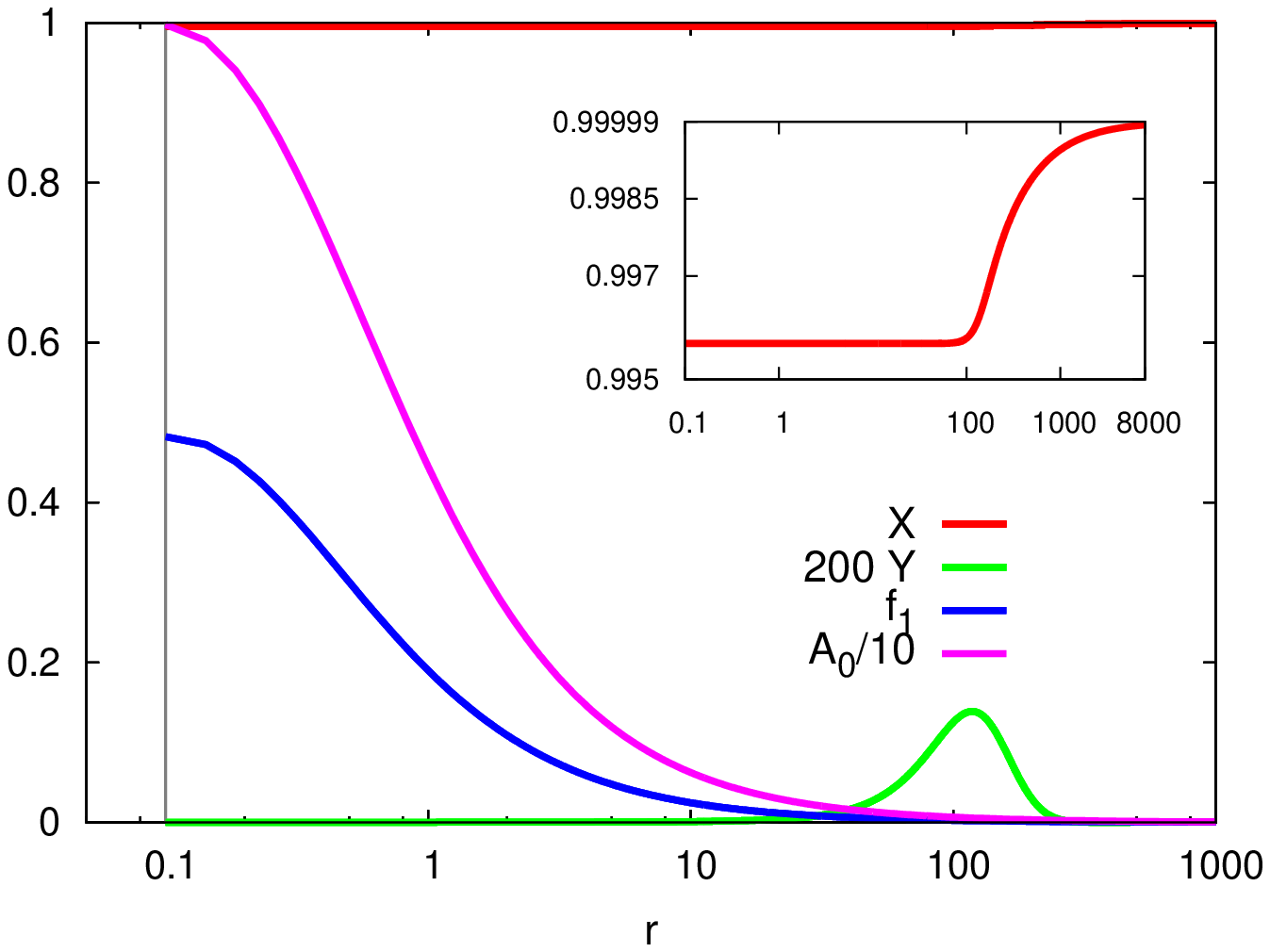}
\includegraphics[height=.34\textheight,  angle =-90]{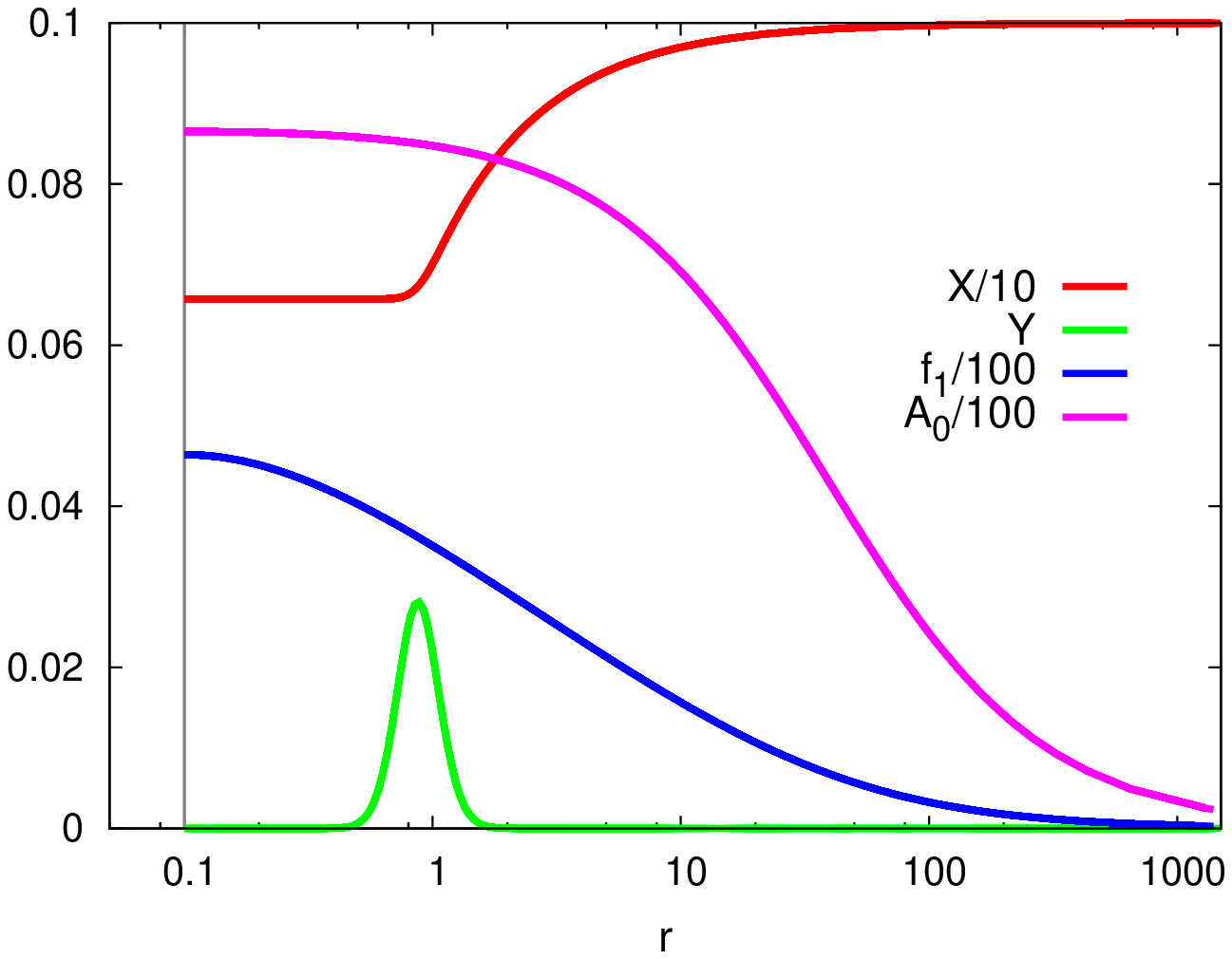}
\includegraphics[height=.34\textheight,  angle =-90]{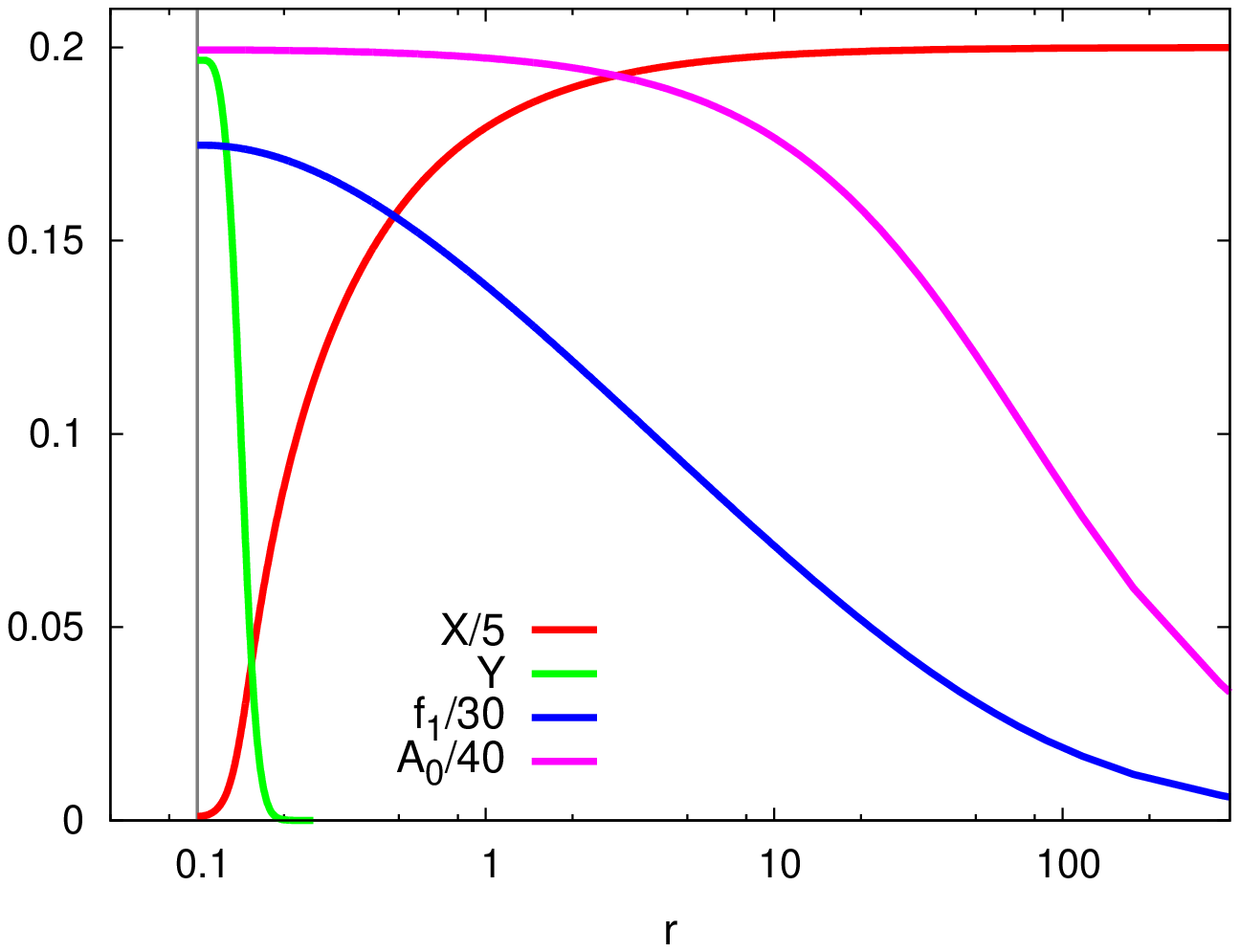}
\includegraphics[height=.34\textheight,  angle =-90]{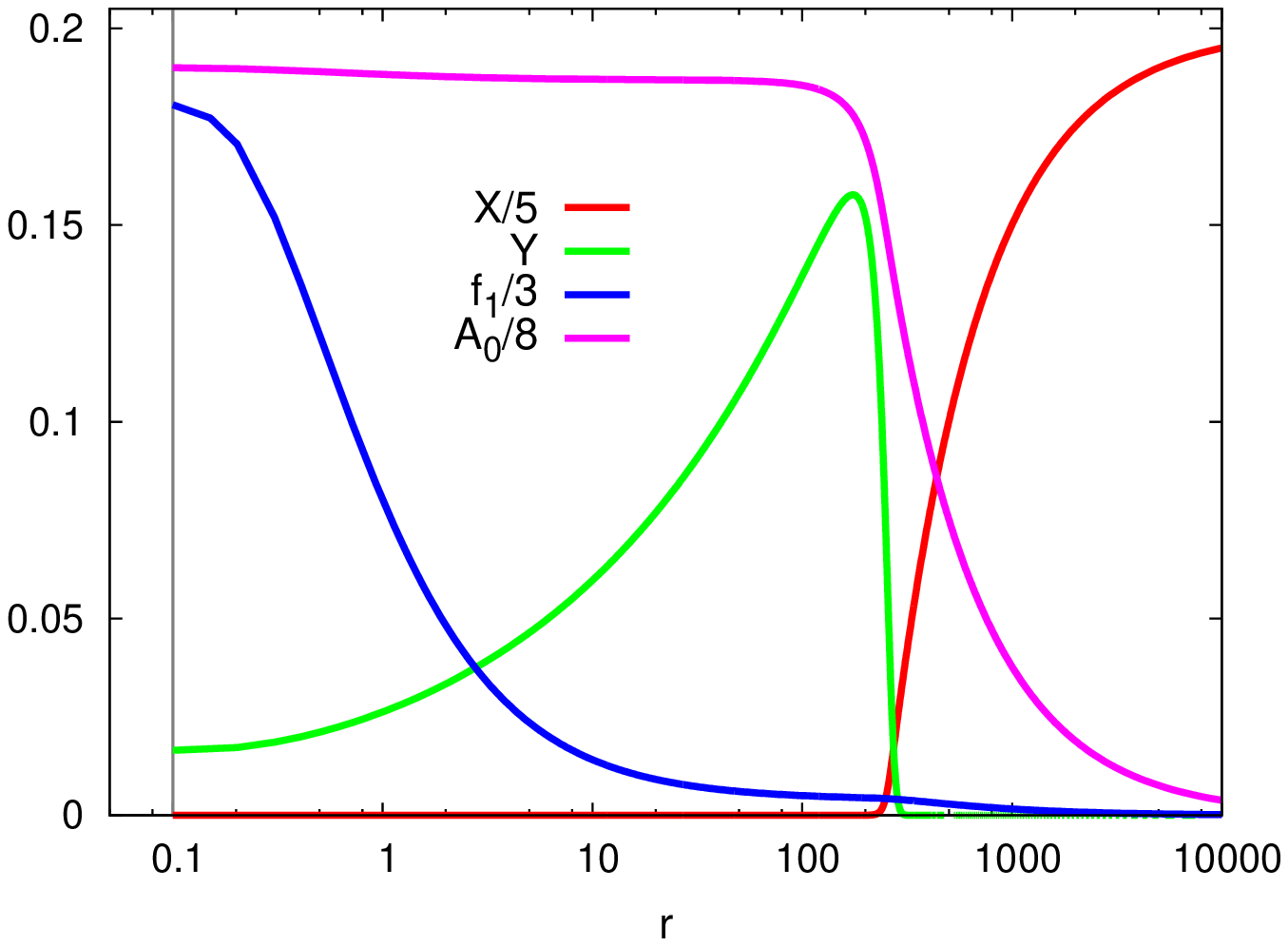}
\end{center}
\caption{\small
Dyonic RN black holes with resonant Q-hair in the massless ($\mu=0$) limit:
The profiles of almost critical solutions approaching the corresponding extremal values of the chemical potential for horizon radius $r_H=0.1$, and gauge coupling $g=0.1$.
The maximal value is approached for chemical potential $\mu_{ch}=9.998$ and gravitational coupling $\alpha=0.10$ (upper left), $\mu_{ch}=8.66$ and $\alpha=0.23$ (upper right), $\mu_{ch}=7.974$ and $\alpha=0.25$ (lower left), and the minimal value for $\mu_{ch}=1.52$ and $\alpha=0.20$ (lower right).}
    \lbfig{fig5}
\end{figure}

Further increase of the coupling $\alpha$ shifts the balance between the electrostatic repulsion and the gravitational attraction that becomes stronger.
Then the branches of hairy dyonic black holes emerge at some maximal value of the chemical potential, which is smaller than the threshold $\mu_{ch}^{(max)} = 10$, as seen in Fig.~\ref{fig3} (upper left).
Indeed, there is an upper limiting value of the chemical potential $\mu_{ch}$ for the dyonic RN black holes, and both the mass $M$ and electric charge $Q$ of the solutions increase to infinity, as this limiting value is approached.
The maximal value $\mu_{ch}^{(max)}$ of the chemical potential decreases as $\alpha$ becomes larger.

In Fig.~\ref{fig5} we display the (scaled) profiles of almost critical solutions approaching their corresponding extremal values of the chemical potential.
The upper left plot corresponds to such a solution for $\alpha=0.1$ with chemical potential $\mu_{ch}=9.998$.
It is analogous to the case illustrated in Fig.~\ref{fig4} for the larger gravitational coupling $\alpha=0.2$, where a small shell formed by the  non-linearly interacting complex doublet emerges in the exterior region of the dyonic RN spacetime.

As $\alpha$ increases further,  still such a small shell emerges, but located closer to the horizon, as seen in Fig.~\ref{fig5} (upper right) for $\alpha=0.23$ and chemical potential $\mu_{ch}=8.66$.
At the same time the real scalar field approaches a 
constant in a small region outside the horizon, and then transitions to a monotonic increase towards its boundary value.
The shell of the complex doublet $\Phi$ is completely localized in the transition region of the real scalar field.
There are two massless fields outside this inner region, the electric potential $A_0$ and the real scalar field $\psi$, both with Coulomb-like asymptotic decay.

This peculiar configuration thus corresponds to a dyonic RN black hole with two types of hair:
The hair formed by the real scalar field $\psi$ consists of an inner ball followed by a Coulomb-like outer decay, whereas the hair formed by the complex scalar doublet $\Phi$ is localized within a thin shell localized around the transition of $\psi$.
Interestingly, neither the metric component $g_{00}$ nor the electric gauge field component $A_0$ seem to reflect this behavior of the scalar fields, but they smoothly decrease towards their asymptotic values.

For still larger $\alpha$ the field of a complex doublet contracts more and more toward the horizon while it decreases in amplitude, as shown in Fig.~\ref{fig5} (lower left).
The real scalar field, on the other hand, decreases more and more steeply in the inner region toward the horizon, while in the outer region it approaches the vacuum value with increasingly smaller slope, i.e., smaller scalar charge $D$.
Indeed, the corresponding scalar charge $D$ of such critical solutions is
very small, as seen in Fig.~\ref{fig3} (lower right).
We conjecture that in the limit we find a step function for the real scalar, and a vanishing complex doublet.
In that case the limiting solution would correspond to a dyonic RN black hole.

Considering the opposite limit of minimal critical value of the chemical potential $\mu_{ch}^{(min)}$, we notice that all global charges of the long-range fields diverge: the ADM mass $M$, the electric charge $Q$  and the scalar charge $D$.
At the same time, the hairiness parameter $h$ approaches unity, as seen in Fig.~\ref{fig3} (middle right).

This limit is illustrated in Fig.~\ref{fig5} (lower right), where we display the (scaled) profile functions of an almost critical solution approaching the minimal value of the chemical potential at $\mu_{ch}\sim 1.49$ for $r_H=0.1$, and coupling $\alpha=0.2$.
The complex doublet is small and fully located inside an inflating bubble.
In contrast, the real scalar field vanishes inside this bubble, and is present only on the outside, where it exhibits a Coulomb-like asymptotic decay.
The electric potential $A_0$, on the other hand, is almost constant in the interior and exhibits the expected Coulomb decay in the outside region.

The current results suggest that toward the approach of the limiting configuration the transition of the functions inside and outside the core region will become increasingly sharp.
The limiting solution is therefore expected to consist of two regions, an inner region, where the electric field and the real scalar field vanish, and an outer region where the complex doublet vanishes, while the electric field and the real scalar exhibit Coulomb-like decay.
Thus in the outer region this solution is expected to correspond to a Penney solution \cite{Penney:1969xk}.
A similar scenario was observed earlier for non-Abelian dyonic solutions \cite{Ibadov:2007qt}.

Since the massless real scalar field $\psi$ becomes zero inside some core region as the minimal value of the chemical potential is approached, the complex doublet $\Phi$ becomes massless there, and the electric field (almost) vanishes in this region.
This shifts the balance between the volume energy of the Q-hair and the surface energy.
The gravitational interaction can no longer stabilize the cloud which rapidly inflates.
The limiting value of the chemical potential $\mu_{ch}^{(min)}$ shows almost no dependence on the strength of the gravitational coupling $\alpha$, as seen in Fig.~\ref{fig3}.
We note, that the minimal critical value of the chemical potential $\mu_{ch}^{(min)}$ is also independent of the horizon area $A_H$.
The electrostatic repulsion between the charged horizon and Q-cloud plays little role as the cloud inflates.

\begin{figure}[t!]
\begin{center}
\includegraphics[height=.34\textheight,  angle =-90]{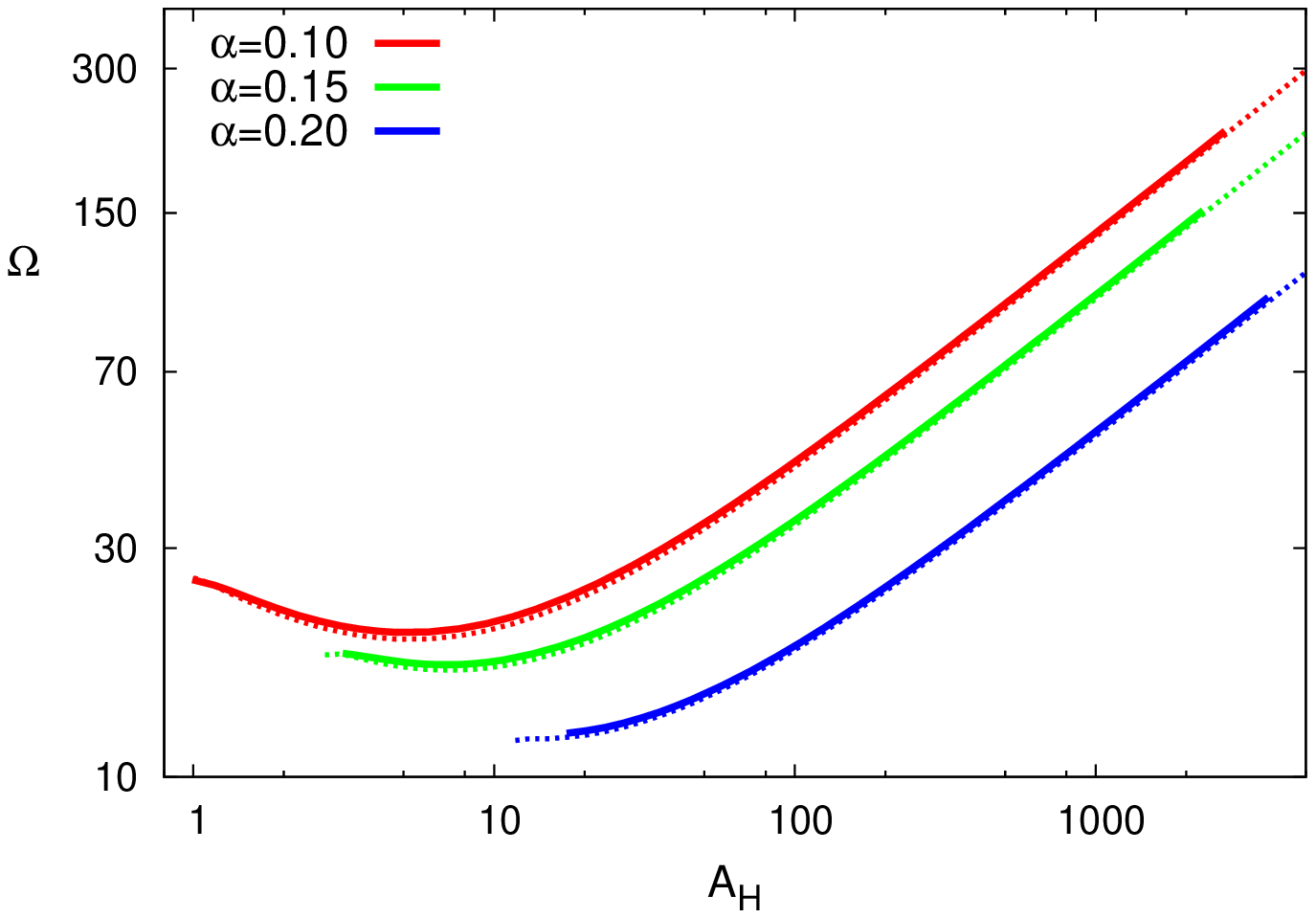}
\includegraphics[height=.34\textheight,  angle =-90]{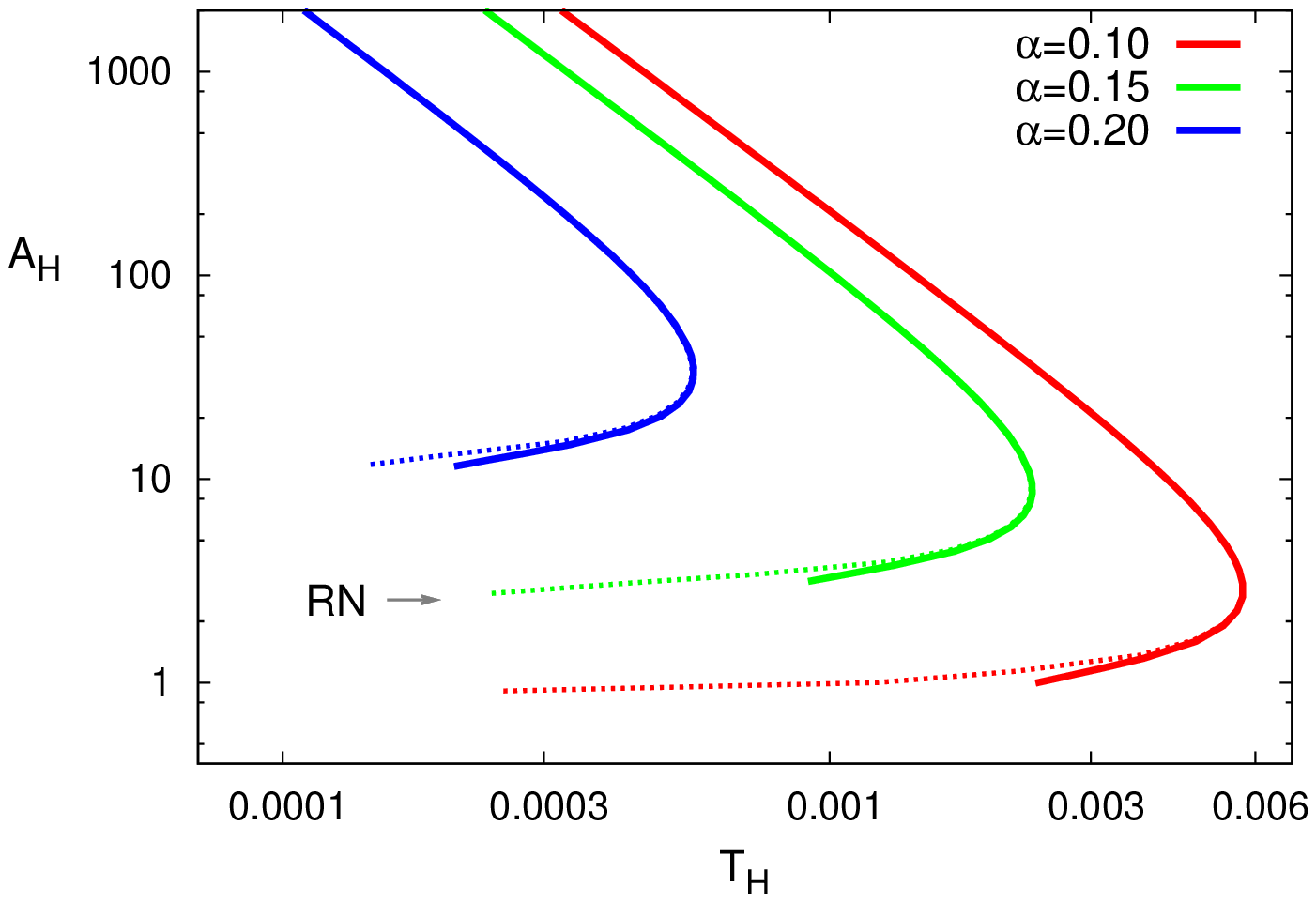}
\end{center}
\caption{\small
Dyonic RN black holes with resonant Q-hair in the massless ($\mu=0$) limit:
Grand potential $\Omega$ (left) and Hawking temperature $T_H$ (right) vs horizon area $A_H$ for
fixed chemical potential $\mu_{ch}=8.5$ and a set of values of the gravitational coupling $\alpha$, for gauge coupling $g=0.1$.
For comparison the corresponding dyonic RN properties are also shown (dotted curves).
}
    \lbfig{fig6}
\end{figure}

Finally, we again address the case where we keep the chemical potential $\mu_{ch}$ fixed and increase the horizon radius $r_H$.
In this case the configuration rapidly evolves towards a limiting solution, that resembles the one discussed above, and corresponds to a hairy dyonic RN black hole with a constant real scalar field in an interior region and a shell-like complex doublet as displayed in Fig.~\ref{fig5} (upper right).
With increasing horizon area $A_H$ the mass $M$ and the scalar charge $D$ of the configurations increase monotonically.

We present in Fig.~\ref{fig6} the dependence of the grand potential $\Omega$ and the Hawking temperature $T_H$ of the solutions on the horizon area $A_H$ (cf. corresponding plots in Fig.~\ref{fig2} for the massive case).
The inset on the left plot illustrates that the mass of the dyonic RN black holes with hair (solid curves) is always slightly higher than the mass of the black holes without hair (dotted curves).

In the massless limit the hairy dyonic RN black holes seem to exist for an arbitrarily large value of the horizon area $A_H$.
This may come as a surprise, since black holes with synchronized hair or resonant hair typically cease to exist at some critical value of the horizon area \cite{Volkov:1998cc,Herdeiro:2015waa}.
This also holds for black holes with Skyrme hair or black holes within magnetic monopoles.
On the other hand, the horizon area of the hairy black holes in Einstein-Yang-Mills theory \cite{Volkov:1989fi} or in the Goldstone model \cite{Radu:2011uj} does not show such a restriction.

\subsection{Dyonic Reissner-Nordstr\"om black holes with scalar clouds}

\begin{figure}[t!]
\begin{center}
\includegraphics[height=.33\textheight,  angle =-90]{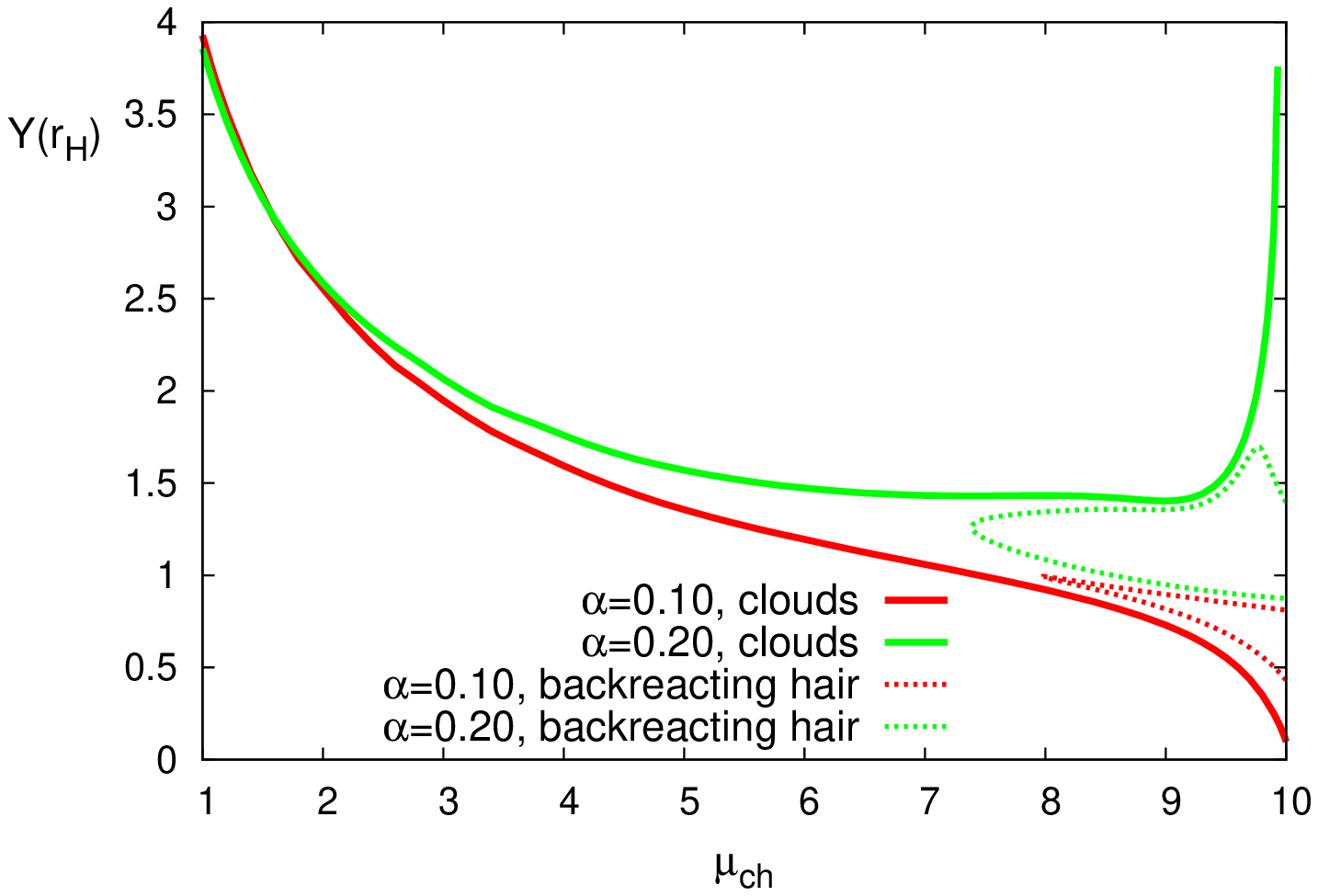}
\includegraphics[height=.33\textheight,  angle =-90]{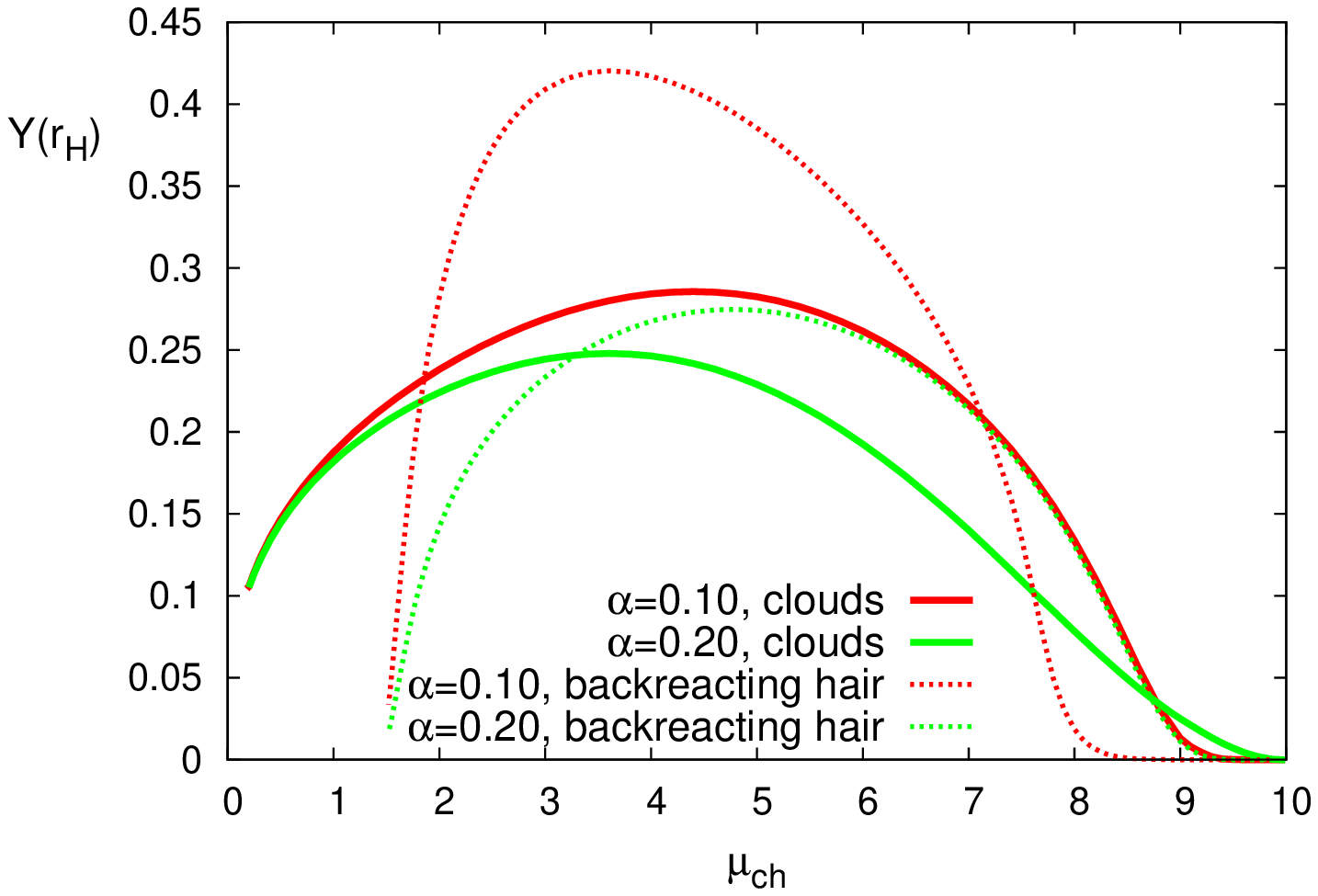}
\end{center}
\caption{\small
Dyonic RN black holes with non-linear FLS-type scalar clouds:
Horizon value of profile function $Y(r_H)$ in the massive case (a) and in the massless ($\mu=0$) limit (b) vs chemical potential $\mu_{ch}$ for two values of the gravitational coupling $\alpha$, for horizon radius $r_H=0.1$, and gauge coupling $g=0.1$.
For comparison the corresponding dyonic RN black holes with backreaction are also shown (dotted curves).
}
\lbfig{fig7}
\end{figure}

We close this section with a discussion of scalar clouds in the dyonic RN background.
We recall that there are no linear clouds in models with a single massive complex scalar field \cite{Hod:2012wmy,Hod:2013nn,Hod:2015hza}.
Likewise, there are no linear clouds for dyonic RN black holes with a self-interacting $U(1)$ gauged scalar doublet \cite{Herdeiro:2024yqa}.
The arguments given in Ref.~\cite{Herdeiro:2024yqa} for the absence of linear clouds carry over to the present case of the EMFLS type model, since in the linear approximation the coupling term of the scalar fields simply reduces to a mass term for the doublet, while the real scalar field will be decoupled.
In contrast to linear clouds, however, non-linear clouds can exist in a dyonic RN background, as we demonstrate in the following.

In the case of a massive scalar field, $\mu>0$, the dyonic RN black holes with non-linear FLS clouds possess the same maximal value $\mu_{ch}^{(max)}$ of the chemical potential as the dyonic hairy RN black holes.
A striking difference is, however, the absence of a minimal value  $\mu_{ch}^{(min)}$ of the chemical potential.
Instead, the FLS-type clouds simply continue for smaller $\mu_{ch}$ along with the dyonic RN black holes.
This is illustrated in Fig.~\ref{fig7}a, where we show the horizon value of profile function $Y(r_H)$ in the massive case versus the chemical potential $\mu_{ch}$ for two values of the gravitational coupling $\alpha$, a horizon radius $r_H=0.1$, and gauge coupling $g=0.1$.
For comparison we have included the corresponding sets of values for the dyonic RN black holes with the scalar backreaction included (dotted curves), as discussed in subsection B.
Figure \ref{fig7}b shows the analogous set of curves in the massless limit $\mu=0$.

\begin{figure}[t!]
\begin{center}
\includegraphics[height=.33\textheight,  angle =-90]{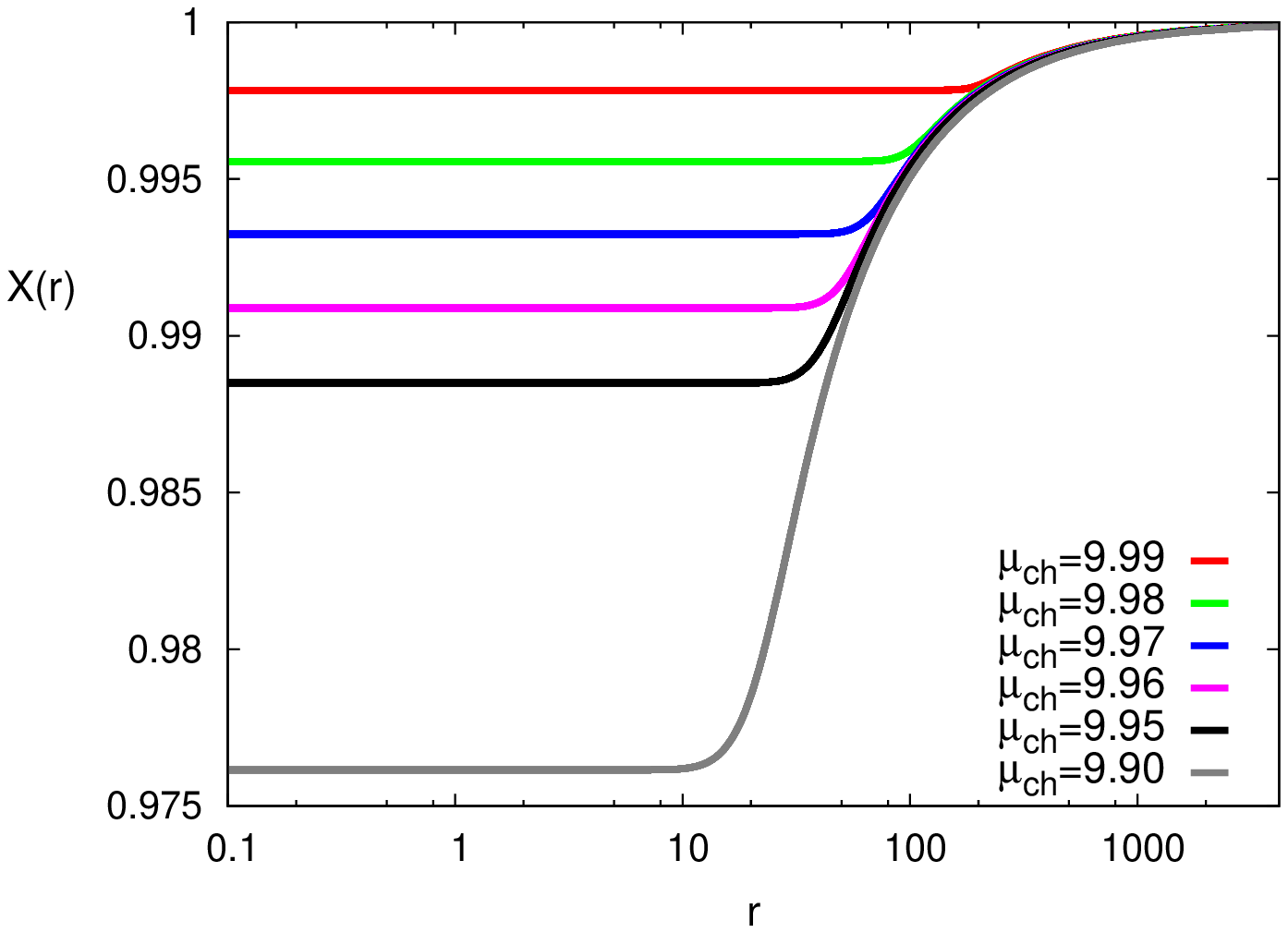}
\includegraphics[height=.33\textheight,  angle =-90]{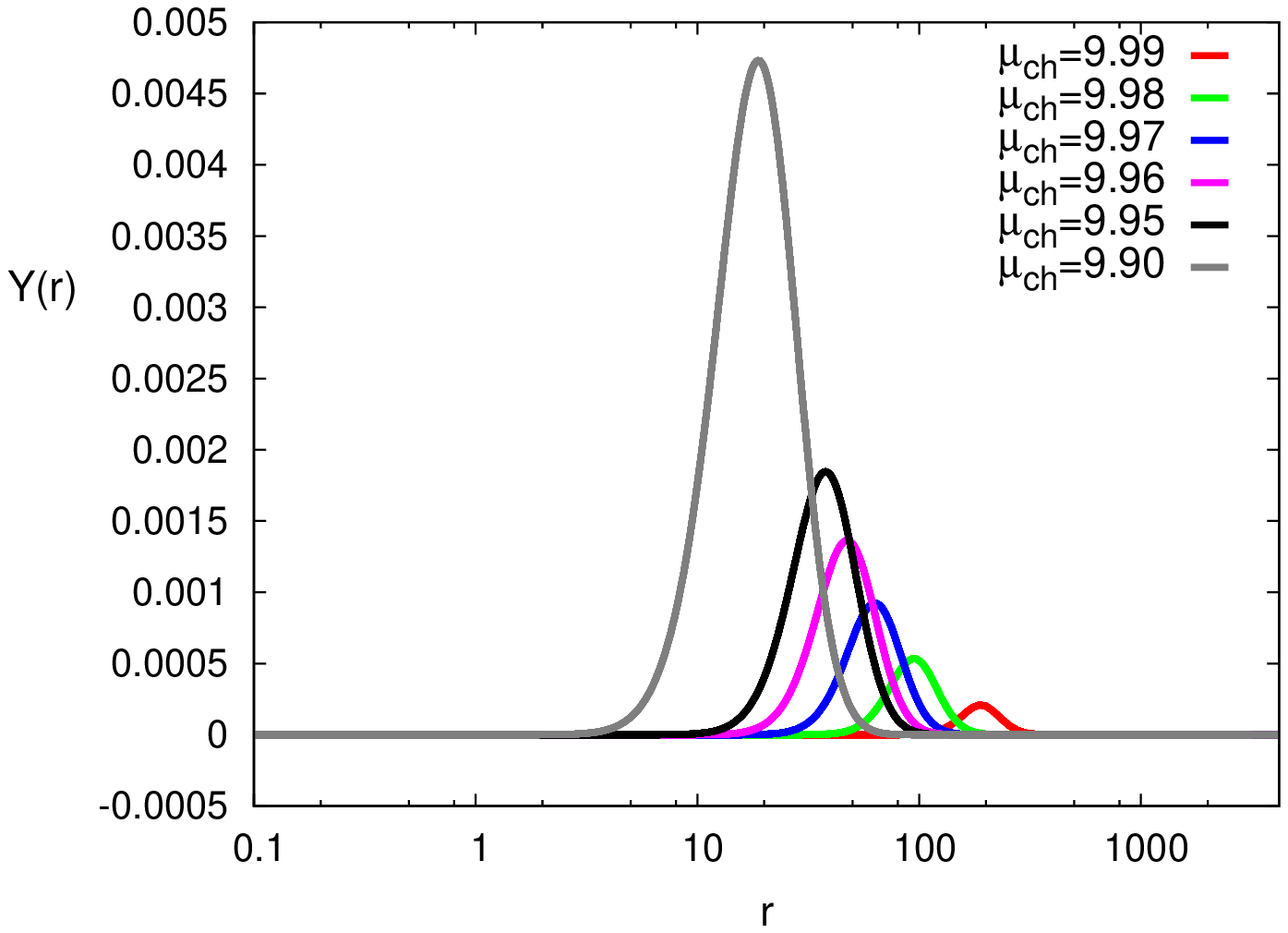}
\end{center}
\caption{\small
Dyonic RN black holes with non-linear FLS-type scalar clouds in the massless ($\mu=0$) limit:
Profile functions of the scalars $X(r)$ (left) and $Y(r)$ (right) for a set of values of the chemical potential $\mu_{ch}$ approaching the maximal value $\mu_{ch}^{(max)}$ of the chemical potential for gravitational coupling $\alpha=0.20$, horizon radius $r_H=0.1$, and gauge coupling $g=0.1$.
}
    \lbfig{fig8}
\end{figure}

In Minkowski space  in the massless limit $\mu=0$, non-linear charged FLS  Q-shells, emerge from the vacuum, when the frequency $\omega$ is decreased below the mass threshold \cite{Loiko:2019gwk}.
Here we observe that similarly, in the background of dyonic RN black holes, non-linear charged FLS-type shells emerge when the chemical potential $\mu_{ch}$ decreases below its maximal value.
Thus the charged scalar doublet forms a shell and not a cloud upon emergence.
We note, that in addition to mimicking the Minkowski case, the FLS-type shells feel the electrostatic repulsion from the horizon.
With decreasing chemical potential $\mu_{ch}$ the attractive forces take over and clouds are formed.

We illustrate the emergence of the non-linear FLS-type shells in the massless ($\mu=0$) limit in Fig~\ref{fig8}, where we exhibit the profile functions of the scalars $X(r)$ (left) and $Y(r)$ (right) for an equidistant set of values of the chemical potential $\mu_{ch}$, that approaches the maximal value $\mu_{ch}^{(max)}$, for gravitational coupling $\alpha=0.20$, horizon radius $r_H=0.1$, and gauge coupling $g=0.1$.
These figures demonstrate strikingly, how the emergence of the non-linear shells proceeds from the vacuum as the chemical potential is decreased from its maximal value.
We note, in particular, the (almost) equidistant changes in the scalar component $X(r)$ with equidistant steps of $\mu_{ch}$, as well as the (almost) linear dependence of the maximum value $Y(r_{max})$ of the shell.


\section*{Conclusions}

In this paper we have discussed static spherically symmetric dyonic RN black holes with resonant hair in a multicomponent EMFLS-type model with symmetry breaking potential.
The matter field sector includes a self-gravitating gauged complex scalar doublet coupled to a real scalar field, that has a finite vacuum expectation value.
Two cases have been considered:~(i) a massive short-ranged real field and (ii) a massless long-ranged real field.
In both cases, Q-hairy RN dyonic black holes are shown to exist, but the presence of magnetic charge excludes a globally regular boson star limit ($r_H\to 0$).

The appearance of dyonic black holes with two types of hair is related to the subtle force balance between the gravitational attraction and the electrostatic repulsion between the horizon of the dyonic RN black hole and the Q-clouds outside the horizon, as well as the forces of the scalar interactions, which may also be long-ranged.

Our results show that for finite values of the mass parameter $\mu$ of the real scalar field, the dyonic black holes with gauged scalar hair share most of the basic properties of the known spherically symmetric hairy RN black holes with electric charge \cite{Herdeiro:2020xmb,Hong:2020miv,Kunz:2021mbm,Herdeiro:2024yqa}.
In this case hairy RN black holes are always disconnected from the corresponding RN  black holes, whose domain of existence is much larger.

The dyonic black holes with Q-hair possess, however, quite different properties when the real scalar field becomes massless.
Depending on the strength of the gravitational coupling, we then observe different scenarios.
As the gravitational coupling remains sufficiently weak, the maximally allowed value of the chemical potential corresponds to the mass threshold.
In such a case a non-linear Q-shell is formed, that arises without a gap in the electro-vac asymptotic region of the dyonic RN black hole.
In fact, when switching off the backreaction of the scalar fields to study Q-clouds in the background of the dyonic RN black hole, a completely analogous pattern is seen.
As the chemical potential is decreased in equidistant steps below its maximal value, where the scalar fields vanish, non-linear Q-shells arise, whose scalar amplitudes increase (almost) linearly with the decreasing chemical potential.

As the gravitational coupling  becomes stronger, the maximal value of the chemical potential decreases below the threshold, and the limiting solution corresponds to an intriguing configuration.
The real scalar field becomes constant within a small inner region outside the horizon and then transitions to a Coulomb-like behavior approaching its vacuum expectation value.
The complex scalar field, on the other hand, becomes small and confined to a shell located in the transition region of the real scalar field.
The ADM mass and further observables then follow closely the corresponding ones of the dyonic RN black holes.
For still larger coupling the complex doublet gets located close to the horizon as the maximal value of the chemical potential is approached, while the real scalar field rises more and more steeply close to the horizon, assuming an almost constant value farther outside.

The minimal value of the chemical potential, on the other hand, is almost independent of the gravitational coupling strength.
In the limit the solution will split into an inner and an outer part as has been observed before for numerous types of hairy black holes.
In the outer part then a well-known solution is obtained, which here should correspond to a Penney solution, since the complex doublet vanishes, while the real scalar rises from zero to its boundary value.

The work here may be taken further by considering axially symmetric gauged Q-hairy dyonic black holes and by including spin.
Another interesting question, we hope to address in the near future, is to investigate higher-dimensional analogues of Q-hairy black holes in a generalized multi-component EMFLS-type model.

\section*{Acknowledgements}
We thank Carlos Herdeiro and Eugen Radu for discussions. Y.S.
would like to thank the Hanse-Wissenschaftskolleg Delmenhorst for support and hospitality.

\begin{small}

\end{small}
\end{document}